\pdfoutput=1
\documentclass[5p,twocolumn]{elsarticle}
\usepackage{graphicx,epstopdf,titlesec,siunitx,amsmath,fixltx2e,subcaption,textgreek,amsmath,lineno,tabularx}
\journal{Icarus}
\biboptions{authoryear}
\newcolumntype{L}{>{\raggedright\arraybackslash}X}

\begin{document}

\begin{frontmatter}

\title{Latitudinal variability in Jupiter's tropospheric disequilibrium species: GeH\textsubscript{4}, AsH\textsubscript{3} and PH\textsubscript{3}}

\author[oxford]{R. S. Giles\corref{cor1}}
\ead{rohini.giles@physics.ox.ac.uk}
\author[leicester]{L. N. Fletcher}
\author[oxford]{P. G. J. Irwin}

\address[oxford]{Atmospheric, Oceanic \& Planetary Physics, Department of Physics, University of Oxford, Clarendon Laboratory, Parks Road, Oxford, OX1 3PU, UK}
\address[leicester]{Department of Physics and Astronomy, University of Leicester, University Road, Leicester, LE1 7RH, UK}

\cortext[cor1]{Corresponding author}

\begin{abstract}

Jupiter's tropospheric composition is studied using high resolution spatially-resolved 5-\textmu m observation from the CRIRES instrument at the Very Large Telescope. The high resolving power (R=96,000) allows us to spectrally resolve the line shapes of individual molecular species in Jupiter's troposphere and, by aligning the slit north-south along Jupiter's central meridian, we are able to search for any latitudinal variability. Despite the high spectral resolution, we find that there are significant degeneracies between the cloud structure and aerosol scattering properties that complicate the retrievals of tropospheric gaseous abundances and limit conclusions on any belt-zone variability. However, we do find evidence for variability between the equatorial regions of the planet and the polar regions. Arsine (AsH\textsubscript{3}) and phosphine (PH\textsubscript{3}) both show an enhancement at high latitudes, while the abundance of germane (GeH\textsubscript{4}) remains approximately constant. These observations contrast with the theoretical predictions from~\citet{wang16} and we discuss the possible explanations for this difference.

\end{abstract}


\end{frontmatter}

%\begin{linenumbers}

%%%%%%%%%%%%%%%%%%%%%%%%%%%%%%%%%%%%%%%%%%%%
%%%%%%%%%%%%%%%%%%%%%%%%%%%%%%%%%%%%%%%%%%%%

\section{Introduction}

The 5-\textmu m atmospheric window is a unique region of Jupiter's spectrum where a dearth of opacity from the planet's principal infrared absorbers gives us access to parts of the atmosphere that are otherwise hidden from view. In the absence of clouds, the sensitivity of the 4.5-5.2 \textmu m spectrum peaks in the 4-8 bar region of Jupiter's atmosphere~\citep{giles15}. These pressure levels correspond to Jupiter's middle troposphere, a region of the planet that is home to many interesting phenomena, including multiple cloud layers, disequilibrium chemical species and dynamics ranging from small-scale meteorological features to planetary-scale overturning.

The atmospheric window was first described by~\citet{gillett69} and the high degree of spatial variability was noted by~\citet{westphal69}. The brightness at 5-\textmu m is very sensitive to the opacity of the tropospheric clouds; in the relatively cloud-free belts, the bright radiation from depth can be observed, but this is partially shielded by the thicker clouds in the zones. Since then, the 5-\textmu m region has been the focus of many studies, from ground-based telescopes, airborne observatories and spacecraft. Ground-based and airborne telescopes generally offer a higher spectral resolution, so these observations are well suited to the study of individual gaseous species in the troposphere. \citet{bjoraker86a} used data from the Kuiper Airborne Observatory to determine the tropospheric composition, and a similar analysis was carried out by~\citet{encrenaz96} using the Infrared Space Observatory. The United Kingdom Infrared Telescope, Canada-France-Hawaii Telescope and the Keck Observatory have been used by several groups in order to target individual gaseous species in discrete regions of the planet~\citep{bezard02,noll89,bjoraker15}.

Germane (GeH\textsubscript{4}), arsine (AsH\textsubscript{3}) and phosphine (PH\textsubscript{3}) are all disequilibrium species in Jupiter's troposphere. Their abundances are determined by temperature-dependent equilibrium reactions and deep in the atmosphere, the temperature is high enough that they are chemically stable. The temperature in the observable troposphere is much lower so we would expect the abundances to rapidly drop off~\citep[e.g.][]{fegley94}. However, GeH\textsubscript{4}, AsH\textsubscript{3} and PH\textsubscript{3} have all been detected in Jupiter's 5-\textmu m window with higher abundances than would be expected from equilibrium chemistry~\citep{fink78,noll89,ridgway76}. Further background about GeH\textsubscript{4}, AsH\textsubscript{3} and PH\textsubscript{3} are given in Sections~\ref{sec:geh4},~\ref{sec:ash3} and~\ref{sec:ph3} respectively.

The apparently enhanced abundances are indicative of rapid vertical mixing, which brings the disequilibrium species upwards from deeper pressures where they are more abundant. Instead of dropping off rapidly with altitude, the abundance is ``frozen in'' at a roughly constant value. This observed abundance corresponds to the equilibrium abundance at a much deeper level, known as the ``quench level", which depends on both the strength of vertical mixing and the chemical kinetic rates~\citep{lewis84}. In the case of PH\textsubscript{3}, the quench level is $\sim$900 K, which corresponds to $>$100 bar~\citep{wang16}. This means that disequilibrium species such as GeH\textsubscript{4}, AsH\textsubscript{3} and PH\textsubscript{3} act as tracers of atmospheric dynamics in Jupiter's troposphere~\citep{taylor04}.

Because vertical mixing strengths vary as a function of latitude~\citep{flasar78,wang15}, the tropospheric abundances of these species are also expected to vary. \citet{wang16} recently carried out a theoretical study where they combined diffusion-kinetics calculations with estimates for how Jupiter's eddy diffusion coefficient varies with latitude, in order to predict the latitudinal distribution of disequilibrium species, including GeH\textsubscript{4}, AsH\textsubscript{3} and PH\textsubscript{3}. Because of the different kinetics rates for the different gases, they predict that the GeH\textsubscript{4} abundance should be a maximum in the equatorial latitudes, and decrease towards the poles, while AsH\textsubscript{3} and PH\textsubscript{3} should be latitudinally uniform.

In this paper, we use high-resolution ground-based observations from the CRIRES instrument at the Very Large Telescope in Chile to search for evidence of latitudinal variability in Jupiter's tropospheric composition. In Sections~\ref{sec:observations} and~\ref{sec:spectral_modelling} we describe our observations and our atmospheric retrieval algorithm. In Sections~\ref{sec:asymmetry} and~\ref{sec:ch3d} we discuss the effect of cloud structure and scattering properties on the atmospheric retrievals. In Sections~\ref{sec:geh4},~\ref{sec:ash3} and~\ref{sec:ph3} we measure the abundances of GeH\textsubscript{4}, AsH\textsubscript{3} and PH\textsubscript{3} respectively. The results of these sections are then discussed in Section~\ref{sec:discussion}.

%%%%%%%%%%%%%%%%%%%%%%%%%%%%%%%%%%%%%%%%%%%%
%%%%%%%%%%%%%%%%%%%%%%%%%%%%%%%%%%%%%%%%%%%%

\section{Observations}
\label{sec:observations}

\subsection{CRIRES observations}

Observations of Jupiter 5-\textmu m window were made using CRIRES, a cryogenic high-resolution infrared echelle spectrograph~\citep{kaufl04}, located at the European Southern Observatory's Very Large Telescope (VLT). The instrument provides long-slit (0.2$\times$40'') spectroscopy across the wavelength range 0.95-5.38 \textmu m, with a resolving power of 96,000. 

The observations were made on 12 November 2012 (05:00-05:40 UT) and 1 January 2013 (02:50-03:30 UT). The atmospheric conditions were better on 12 November 2012, so the majority of this work focuses on those observations. However, the observations from 1 January 2013 are used in Sections~\ref{sec:ash3} and~\ref{sec:ph3} to show that the same conclusions can be drawn from both datasets.

\begin{figure}
\centering
\includegraphics[width=7cm]{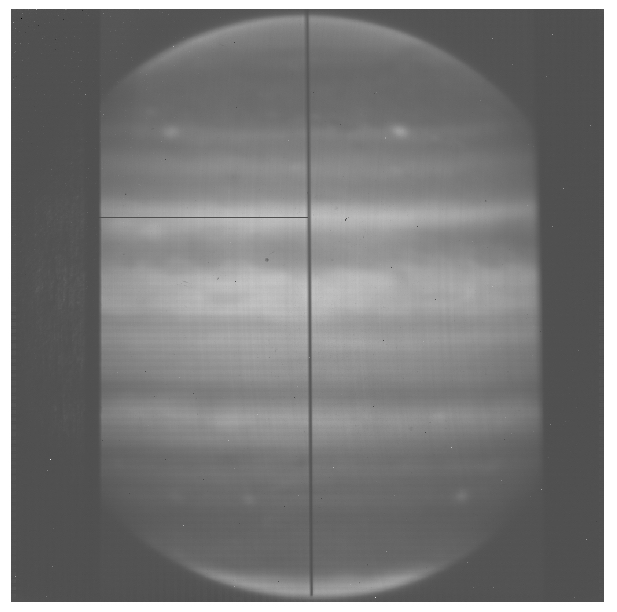}
\caption{Acquisition image of Jupiter made at 05:02 UT on 12 November 2012, immediately before the 5-\textmu m observations. This image was made using a Ks-band filter (2.2 \textmu m). The vertical black line shows the alignment of the narrow slit.}
\label{fig:acquisition}
\end{figure}

\begin{figure}
\centering
\includegraphics[width=9.cm]{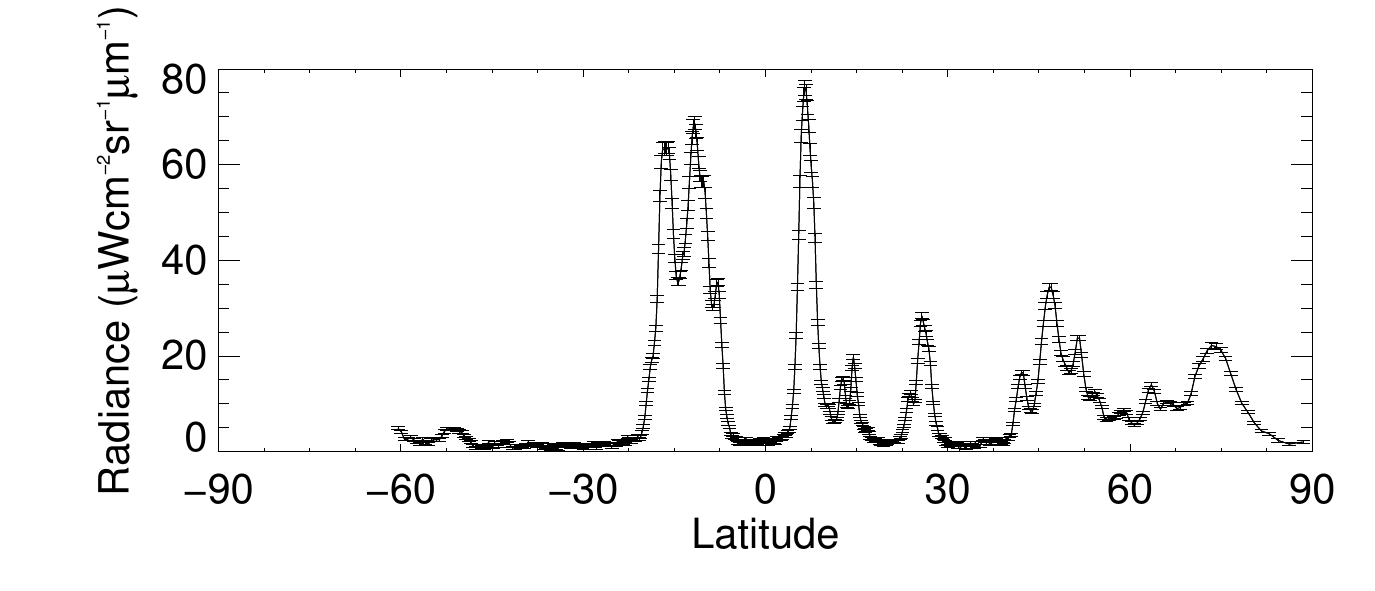}
\caption{Measured radiance at 5.0 \textmu m as a function of latitude.}
\label{fig:latitudinal_profile}
\end{figure}

On both dates, the slit was aligned north-south on Jupiter, along the planet's central meridian, allowing us to measure spatial variability with latitude, but not longitude. Figure~\ref{fig:acquisition} shows the acquisition image of Jupiter made on 12 November, using a Ks-band filter (2.2 \textmu m). The vertical black line shows how the narrow slit was aligned. The slit length (40") was smaller than the angular diameter of Jupiter on the observation dates (47-48"), so the observations cut off the southern polar region. Figure~\ref{fig:latitudinal_profile} shows the observed radiance as a function of latitude at 5.0 \textmu m. The high degree of spatial variability between the belts and the zones is due to the varying opacity of the tropospheric clouds.

On each date, observations were made in 14 different wavelength settings, which together cover the entire 4.55-5.26 \textmu m wavelength range. The time lag between the first and last observation was $\sim$40 minutes; during this time period, Jupiter rotated $\sim$25$^{\circ}$, so the 14 different observations of different spectral regions do not correspond to precisely the same locations on the planet.

Observations were made of a standard star in order to provide radiometric calibration. Pi-2 Orionis (HIP 22509) and Zeta Tauri (HIP 26451) were used for the November and January observations respectively. Dark images and flatfields were also taken.

\subsection{Data reduction}

Initial data reduction was performed using EsoRex, the ESO Recipe Execution Tool~\citep{ballester06}. The flatfields, darks, and nodded pairs of observations were used to produce a single combined observation of Jupiter for each of the 14 wavelength segments. The same process was applied to the observations of the standard star. In both cases, wavelength calibration was achieved by comparing telluric absorption lines with the HITRAN line database~\citep{rothman09}. EsoRex was also used to calculate the background noise on both the stellar and the planetary observations.

Subsequent data reduction was performed independently of the EsoRex software. This included straightening the observations, to correct observed shears in both the spectral and spatial directions. The spectral shear was corrected using the wavelength calibration map generated by EsoRex. The spatial shear was cross-correlating by cross-correlating the north-south latitudinal profiles at different wavelengths i.e. requiring the belts and zones to be located at the same pixel values for all wavelengths. The jovian spectra were divided through by the standard star observations in order to remove the telluric lines, and were multiplied by the standard star reference spectrum to convert Jupiter's photon count into spectral radiance. For Pi-2 Orionis, the standard star reference spectrum was provided by EsoRex. For Zeta Tauri, a black-body curve was scaled to match the observed magnitudes from the 2MASS all-sky survey~\citep{cutri03}. Because the slit is very narrow, some stellar flux will be lost; this was accounted for in the radiometric calibration by assuming that the star is centrally placed in the slit, using a measurement of the star's point spread function made in the perpendicular direction. Once this calibration has been applied, we find that the 5-\textmu m brightness temperatures vary between $\sim$190 K in the coolest parts of the planet and $\sim$260 K in the brightest regions (see Figure~\ref{fig:latitudinal_profile}), which is roughly consistent with the 180-240 K range found in the Cassini VIMS 5-\textmu m observations~\citep{giles15}. The noise on the spectra is roughly 5-10\%, but this does not include the systematic error that could be introduced by the assumptions in the radiometric calibration. These systematic errors could increase or decrease the average radiance, but will not alter the shape of the spectra. The average radiance is primarily determined by the opacity of the upper cloud deck~\citep{giles15}. Uncertainties in the radiometric calibration can therefore alter the apparent thickness of the clouds, but will not affect the analysis of the molecular species. 

Geometric data was calculated using information from JPL HORIZONS, and by assuming that the slit was aligned with the central meridian. Planetocentric latitudes were assigned to each pixel using information about Jupiter's angular size at the observation time, the angular size of each pixel, the centre-of-target latitude, and the location of the planet's limb. In addition, the emission angle, solar zenith angle and azimuthal angle were calculated for each pixel. 

Due to Jupiter's rotation, each of the 14 wavelength segments observed a slightly different longitude, with a $\sim25^{\circ}$ difference between the first and last segment. Because Jupiter is longitudinally inhomogeneous, this means that each wavelength segment is observing a different cloud opacity; this prevents the different segments from being joined together into a single continuous observation. These latitudinally resolved spectra will be used in the following sections in order to search for variability in the deep cloud structure (Section~\ref{sec:ch3d}) and in the abundances of germane, arsine and phosphine (Sections~\ref{sec:geh4}-\ref{sec:ph3}).

%%%%%%%%%%%%%%%%%%%%%%%%%%%%%%%%%%%%%%%%%%%%
%%%%%%%%%%%%%%%%%%%%%%%%%%%%%%%%%%%%%%%%%%%%

\section{Spectral modelling}
\label{sec:spectral_modelling}

The CRIRES spectra were analysed with the NEMESIS retrieval algorithm developed by~\citet{irwin08}. NEMESIS uses a radiative transfer code to compute the top-of-atmosphere spectral radiance for a given atmospheric profile, and then iteratively adjusts the atmospheric parameters in order to produce an optimal fit to the observed spectrum. For this work, we use line-by-line calculations in the radiative transfer code; previous work has made use of the correlated-k approximation~\citep[e.g.][]{giles15} but the high spectral resolution of the CRIRES observations means that this approximation would not save significant computational time. We therefore use the more accurate line-by-line method. The majority of the spectral line data were obtained from the HITRAN 2008~\citep{rothman09} or GEISA 2009~\citep{jacquinet-husson11} molecular databases. The two exceptions are AsH\textsubscript{3}, where we use laboratory data from~\citet{dana93} and~\citet{mandin95}, and GeH\textsubscript{4}, where we use the Spherical Top Data System~\citep[see Section~\ref{sec:geh4_introduction} for more details]{wenger98}. Collision-induced absorption data is taken from~\citet{borysow86},~\citet{borysow87},~\citet{borysow88} and~\citet{borysow91}. We add a conservative 5\% forward modelling error to the retrieval calculations, to take into account inaccuracies in the model (e.g. line data).

NEMESIS has the ability to model a multiple scattering atmosphere using the plane-parallel matrix operator method of~\citet{plass73}, and we show in Section~\ref{sec:asymmetry} that the scattering properties of Jupiter's clouds have significant effects on the retrieved gaseous abundances. Unless otherwise stated, we use the simple cloud model from~\citet{giles15}: a single, spectrally-flat, compact cloud layer, located at 0.8 bar, with a single-scattering albedo $\omega=0.9$ and a Henyey-Greenstein asymmetry parameter $g=0.8$.

The reference atmospheric profile is described as in~\citet{giles15}. The gaseous species with spectral features in the 5-\textmu m region are NH\textsubscript{3}, PH\textsubscript{3}, CO, GeH\textsubscript{4}, AsH\textsubscript{3}, H\textsubscript{2}O and CH\textsubscript{3}D. Unless otherwise stated, the abundances of each of these gases are allowed to vary via a single scaling parameter. 

%%%%%%%%%%%%%%%%%%%%%%%%%%%%%%%%%%%%%%%%%%%%
%%%%%%%%%%%%%%%%%%%%%%%%%%%%%%%%%%%%%%%%%%%%

\section{Scattering properties of upper cloud}
\label{sec:asymmetry}

The high spectral resolution of the CRIRES instrument allows us to resolve the individual absorption lines of molecular species in Jupiter's troposphere. Using the NEMESIS retrieval algorithm, we can retrieve the abundances of gases, and map how these abundances vary across the planet. The high spectral resolution helps to reduce the degeneracy between different gaseous species. However, there is a degeneracy between the gaseous abundances and scattering parameters of Jupiter's tropospheric clouds: specifically, the asymmetry parameter, $g$, which defines the extent to which the cloud particles are forward scattering. 

\begin{figure*}
\centering
\includegraphics[width=12cm]{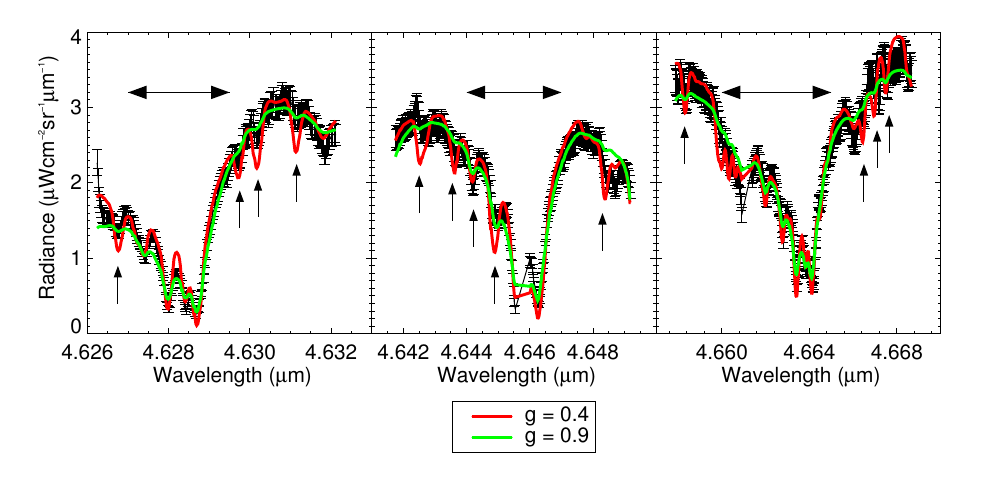}
\caption{Fits to three CH\textsubscript{3}D absorption features using two different values of the cloud particle asymmetry parameter, g. The observational data (black) is from the Equatorial Zone of Jupiter. The location of the CH\textsubscript{3}D absorption features are marked by the horizontal double-headed arrows. The location of the most prominent Fraunhofer lines are marked by the vertical single-headed arrows.} 
\label{fig:fraunhofer_ch3d}
\end{figure*}

An example of this effect is shown in Figure~\ref{fig:fraunhofer_ch3d}. This figure shows three prominent CH\textsubscript{3}D absorption features, marked by the horizontal double-headed arrows. The CRIRES data are shown in black, and is from the Equatorial Zone of the planet. The Equatorial Zone was chosen as it is one of the cloudiest regions of the planet. Small segments of data with high error bars (due to telluric contamination) have been removed. The three absorption features were fitted simultaneously, and the coloured lines show the fits that can be obtained using two different assumptions about the scattering properties of the single cloud deck. Both cases can produce a good fit to the broad CH\textsubscript{3}D features, but the retrieved CH\textsubscript{3}D abundances are significantly different: 0.16 ppm for the case where $g=0.4$ and 0.08 ppm for the case where $g=0.9$.

This effect is due to the different amount of reflected sunlight in the two models. A higher asymmetry parameter (more forward scattering) leads to a higher fraction of radiation originating from the planet, and a lower fraction from reflected sunlight. In contrast, a lower asymmetry parameter (less forward scattering) leads to a higher fraction due to reflected sunlight. A higher $g$ value therefore leads to stronger absorption features for a given gaseous abundance, and hence lower retrieved abundances. It should be noted that this is dependent on the vertical location of the cloud; if the cloud were located below the main line forming region, the effect would be less pronounced. However,~\citet{giles15} showed that the main tropospheric cloud deck must be located well above the line forming in order to account for the highly variable 5-\textmu m radiation.
%0.155, 0.078

This degeneracy between cloud scattering properties and retrieved gaseous abundances is only present in regions of the planet with thick cloud cover (zones like the Equatorial Zone). The belts of the planet are relatively cloud-free, so changing the cloud parameters has little impact on the retrievals. This means that different assumptions about the cloud properties can affect the latitudinal profiles of the gases; one set of parameters could suggest that a molecular species has a constant abundance across the planet, while another set of parameters could suggest that the abundance varies between belts and zones. Selecting the correct scattering parameters is therefore of critical importance.

In~\citet{giles15} we constrained the scattering parameters of Jupiter's tropospheric clouds using limb darkening observations from the VIMS instrument on the Cassini spacecraft. This analysis ruled out strongly forward scattering particles ($g>0.9$), but a broad range of values were consistent with the data. In order to further constrain the asymmetry parameter properties of the clouds, we can make use of the Fraunhofer lines that are present in the CRIRES data because of the reflected sunlight. These narrow solar absorption lines can be seen in Figure~\ref{fig:fraunhofer_ch3d} (marked by the vertical single-headed arrows). The strength of these features also depends on the scattering properties: lower $g$ values lead to higher fractions of reflected sunlight and deeper absorption features.~\citet{bjoraker15} also made use of a Fraunhofer line at 4.67 \textmu m to determine the reflectance of the upper tropospheric clouds.

We ran retrievals using a range of different $g$ values in order to determine which value gave the best fit to the CRIRES data from the EqZ. The entire 5-\textmu m spectral range was used, excluding regions with high error bars due to telluric contamination. For each value of $g$, the goodness-of-fit was calculated and the results are given in Table~\ref{tab:asymmetry}. The best fit was obtained for $g=0.8$, which is consistent with previous analyses~\citep{giles15,irwin08}. In the following sections, we will use a cloud with an asymmetry parameter $g=0.8$. We also make the assumption that the asymmetry parameter does not vary with latitude. In each case, we will investigate the extent to which the results are dependent on the assumed asymmetry parameter.

\begin{table}
\centering
\begin{tabular}{ l | l }
g & \textchi\textsuperscript{2}/N \\ \hline
0.40 & 2.25 \\ 
0.45 & 2.11 \\ 
0.50 & 1.95 \\
0.55 & 1.80 \\
0.60 & 1.66 \\
0.65 & 1.54 \\
0.70 & 1.43 \\
0.75 & 1.31 \\
\textbf{0.80} & \textbf{1.19} \\
0.85 & 1.52 \\
0.90 & 2.40 \\
\end{tabular}
\caption{The goodness-of-fit values (\textchi\textsuperscript{2}/N) for a range of different asymmetry parameters. \textchi\textsuperscript{2}/N is minimised when $g=0.8$.}
\label{tab:asymmetry}
\end{table}

%%%%%%%%%%%%%%%%%%%%%%%%%%%%%%%%%%%%%%%%%%%%%
%%%%%%%%%%%%%%%%%%%%%%%%%%%%%%%%%%%%%%%%%%%%%

\section{CH\textsubscript{3}D and the deep cloud structure}
\label{sec:ch3d}

\subsection{Introduction}

Methane (CH\textsubscript{4}) was one of the first gases to be discovered in Jupiter's atmosphere when~\citet{wildt32} identified previously unknown spectral features as methane and ammonia absorption lines. It is the third most abundant gaseous species in Jupiter's upper atmosphere, with a volume mixing ratio (VMR) of $2.04\times10^{-3}$ in the troposphere detected by the Galileo Probe Mass Spectrometer~\citep{wong04b}. Unlike many other gaseous species, this abundance is not expected to have any spatial variability in the troposphere. This is because CH\textsubscript{4} is chemically stable throughout the troposphere and does not condense at the temperatures found in Jupiter's atmosphere~\citep{taylor04}, which means that it should be well-mixed.

As a methane isotopologue,  CH\textsubscript{3}D (deuterated methane) is also constant throughout Jupiter's troposphere. Unlike CH\textsubscript{4}, no in situ measurements have been made of the jovian CH\textsubscript{3}D abundance, but a tropospheric abundance of 0.16$\pm$0.05 ppm has been estimated from observations made with the Infrared Space Observatory~\citep{lellouch01}. The CH\textsubscript{3}D abundance is particularly relevant to our study because CH\textsubscript{3}D has several strong absorption features in the 5-\textmu m window. 

It is the spatial homogeneity of CH\textsubscript{3}D that provides an opportunity to study Jupiter's deep cloud structure. Because the CH\textsubscript{3}D abundance is expected to be constant with latitude, any observed variation in the spectral lineshape must be due to another factor. One way of reproducing variable lineshapes is by varying the deep tropospheric cloud structure~\citep{bjoraker15}. In this section, we will search for spatial variability in the CH\textsubscript{3}D lineshapes, and will explore how this can be interpreted in terms of cloud structure.

\subsection{Analysis}

\subsubsection{CH\textsubscript{3}D}
\label{sec:ch3danalysis}

\begin{figure*}
\centering
\includegraphics[width=12cm]{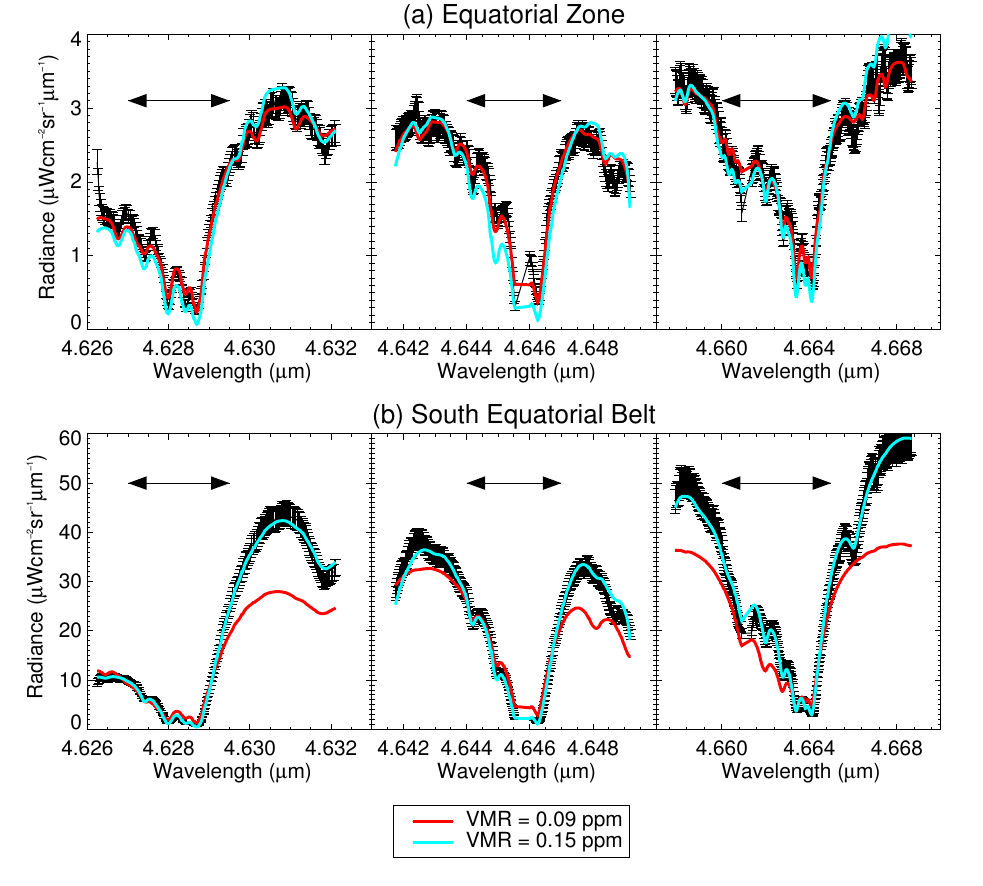}
\caption{CH\textsubscript{3}D absorption feature fits in two different regions of the planet. CRIRES data is shown in black, and the CH\textsubscript{3}D features are shown by the horizontal arrows. Each colour shows the fit obtained using different CH\textsubscript{3}D abundances: 0.09 ppm (the best fit value for the EqZ) and 0.15 ppm (the best fit value for the SEB).} 
\label{fig:ch3d_fits}
\end{figure*}

There are several CH\textsubscript{3}D features in the 5-\textmu m window. Three of the strongest features are shown in Figure~\ref{fig:ch3d_fits}. These three features are within the same wavelength segment, which means they are from the same longitude (i.e. were observed simultaneously) and can therefore be analysed simultaneously. The CRIRES data are shown in black, and the arrows show the locations of the CH\textsubscript{3}D features. The upper panel shows the data from the cool Equatorial Zone (EqZ, 5$^{\circ}$S-4$^{\circ}$N), while the lower panel shows data from the warm South Equatorial Belt (SEB, 16$^{\circ}$S-6$^{\circ}$S). Figure~\ref{fig:ch3d_fits} shows that the CH\textsubscript{3}D features are considerably narrower in the EqZ than they are in the SEB. 

This difference was further explored by performing retrievals with the NEMESIS retrieval algorithm. Based on the conclusion of Section~\ref{sec:asymmetry}, we initially assumed a cloud structure made up of a single compact cloud layer at 0.8 bar, with scattering properties \textomega\:= 0.9 and g = 0.8. The optical thickness of this cloud was allowed to vary, along with the abundances of the molecular species as described in Section~\ref{sec:spectral_modelling}. For the SEB, the retrieved CH\textsubscript{3}D abundance was 0.15 ppm, approximately the same as the value obtained by~\citet{lellouch01}. However, the EqZ produced a much lower value of 0.09 ppm.

As a further test, additional retrievals were run where the EqZ CH\textsubscript{3}D abundance was fixed to the higher level and the SEB abundance was fixed to the lower level. The results of this are shown by the coloured lines in Figure~\ref{fig:ch3d_fits} and they show that the EqZ and the SEB cannot be fit using the same CH\textsubscript{3}D abundance. A VMR of 0.15 ppm fits the SEB data, but is too broad for the EqZ data, and a VMR of 0.09 ppm fits the EqZ data but is too narrow for the SEB.

However, a factor $\sim$2 difference in CH\textsubscript{3}D abundance for different regions of the planet is not physically plausible. As described in the previous section, methane and its isotopologues are expected to be well-mixed in the troposphere, with no spatial variability. We must therefore consider an alternative explanation for this change in CH\textsubscript{3}D lineshape.

\subsubsection{Deep cloud structure}
\label{sec:deepcloudstructure}

One possibility is spatial variability in the deep cloud structure. As described by~\citet{bjoraker15}, this can be modelled by considering the presence of deep clouds at the pressure levels where water is expected to condense ($\sim$5 bar). If there is very little cloud opacity at these pressures, the radiation is originating from deep in the atmosphere, where the pressure is higher. Pressure broadening therefore leads to a lineshape with broad wings. Inserting an optically thick cloud layer at 5 bar has two effects on the spectrum: (i) it reduces the continuum radiance, decreasing the apparent depth of the absorption feature and (ii) it moves the weighting function upwards to a region of lower pressures and cooler temperatures, leading to a narrower lineshape.

\begin{figure*}
\centering
\includegraphics[width=12cm]{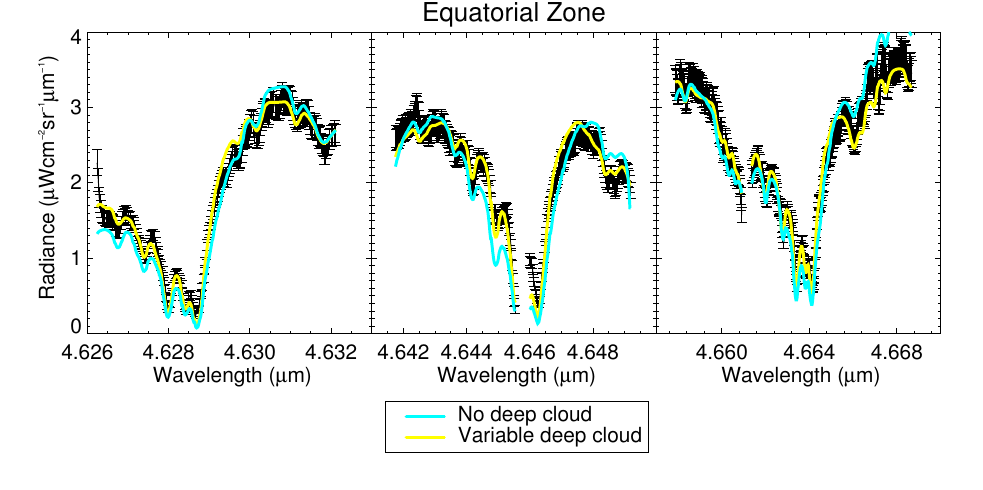}
\caption{Retrievals of three CH\textsubscript{3}D features in the cool EqZ, with varying the cloud structure. The blue line shows the best fit when there is no deep cloud present, and the yellow line shows the fit where the opacity of a deep cloud at 5 bar is allowed to vary. In both cases, the CH\textsubscript{3}D abundance is fixed at 0.16 ppm.}
\label{fig:ch3d_deepcloud}
\end{figure*}

Since the presence of a deep cloud can act to narrow the lineshape, this suggests that we may be able to fit the narrow EqZ absorption features while keeping the CH\textsubscript{3}D abundance fixed at the retrieved value for the SEB. To test this we inserted a deep cloud deck at 5 bar. We assumed the same simple properties as the upper cloud: compact, spectrally flat, \textomega\:= 0.9 and g = 0.8. We then performed a retrieval where the opacity of this deep cloud was allowed to vary. The CH\textsubscript{3}D abundance was held fixed at 0.149 ppm and, as before, the upper cloud and other gaseous species were also allowed to vary. The results are shown by the yellow line in Figure~\ref{fig:ch3d_deepcloud}. For comparison, we also show the case where there is no deep cloud in blue (identical to the blue line in Figure~\ref{fig:ch3d_fits}a). 

The comparison between the blue and yellow lines shows that this addition considerably improves the fit. The retrieved deep cloud opacity  in the EqZ is 230, making the cloud essentially opaque. This agrees with~\citet{bjoraker15}, where the deep cloud in the EqZ was assumed to be opaque. As with~\citet{bjoraker15}, this shows that the difference in CH\textsubscript{3}D lineshape between the EqZ and the SEB can be accounted for by the variations in the deep cloud structure; the deep cloud is transparent in the SEB and opaque in the EqZ.

Based on~\citet{bjoraker15}, the initial analysis was performed using an opaque cloud at 5 bar. We carried out further retrievals in order to test the sensitivity to the cloud location. If the cloud is placed too deep, it falls outside the pressure range probed by 5-\textmu m spectra, and can no longer influence the spectral shape. If the cloud is placed too high, it simply plays the same role as the original 0.8-bar cloud and so does not improve the fit. However, between these two extremes, we find that a broad range of cloud locations from 2 to 7 bar can provide a good fit to the data, with deeper pressure levels requiring higher cloud opacities. In the absence of further constraint, we continue the analysis using the original 5 bar value.

\subsubsection{Latitudinal retrieval}

\begin{figure}
\centering
\includegraphics[width=9cm]{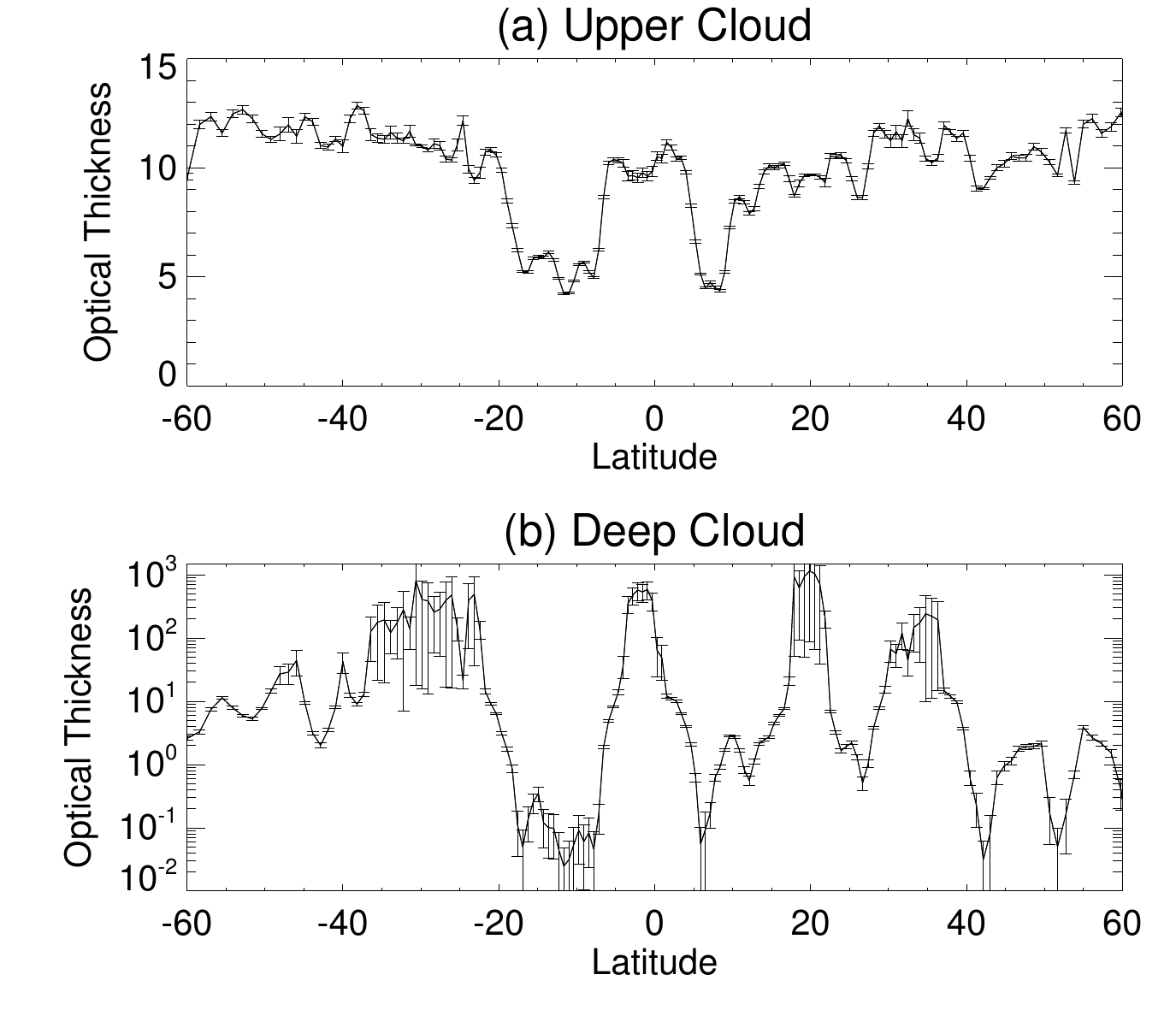}
\caption{Retrieved cloud optical thicknesses as a function of latitude. The retrievals show that the cool zones are fit using an opaque deep cloud, while the warm belts are fit using an optically thin deep cloud.}
\label{fig:cloud_latitude}
\end{figure}

In the previous sections, we were just comparing two discrete parts of the planet, the EqZ and the SEB. We now extend that analysis to perform a full latitudinal retrieval of the deep cloud opacity, using the same CH\textsubscript{3}D features as in the previous sections. Spectra were smoothed in the spatial direction with a width of 5 pixels in order to reduce the noise, and the sampling rate was 3 pixels. For each smoothed spectrum, we ran a retrieval following the same procedure as in Section~\ref{sec:deepcloudstructure}. The retrieved cloud optical thicknesses are shown in Figure~\ref{fig:cloud_latitude}. In the belts, the retrieved deep cloud opacity is $\sim$0.1 (relatively transparent), and for the zones the retrieved deep cloud opacity is $\sim$100 (relatively opaque).

The absolute values of the both cloud optical thickness are very dependent on (i) the accuracy of the radiometric calibration, (ii) the cloud locations and (iii) the choice of cloud scattering parameters. The error bars shown in Figure~\ref{fig:cloud_latitude} do not include these error sources. Changes to either of the first two factors result in a cloud opacity profile that is either scaled up or scaled down, but the relative shape of the latitudinal profile remains constant. The third factor can affect the relative shape of the latitudinal profile, as described in Section~\ref{sec:ch3d_gsensitivity}.

\subsubsection{Sensitivity to the scattering parameters}
\label{sec:ch3d_gsensitivity}

\begin{figure}
\centering
\includegraphics[width=9cm]{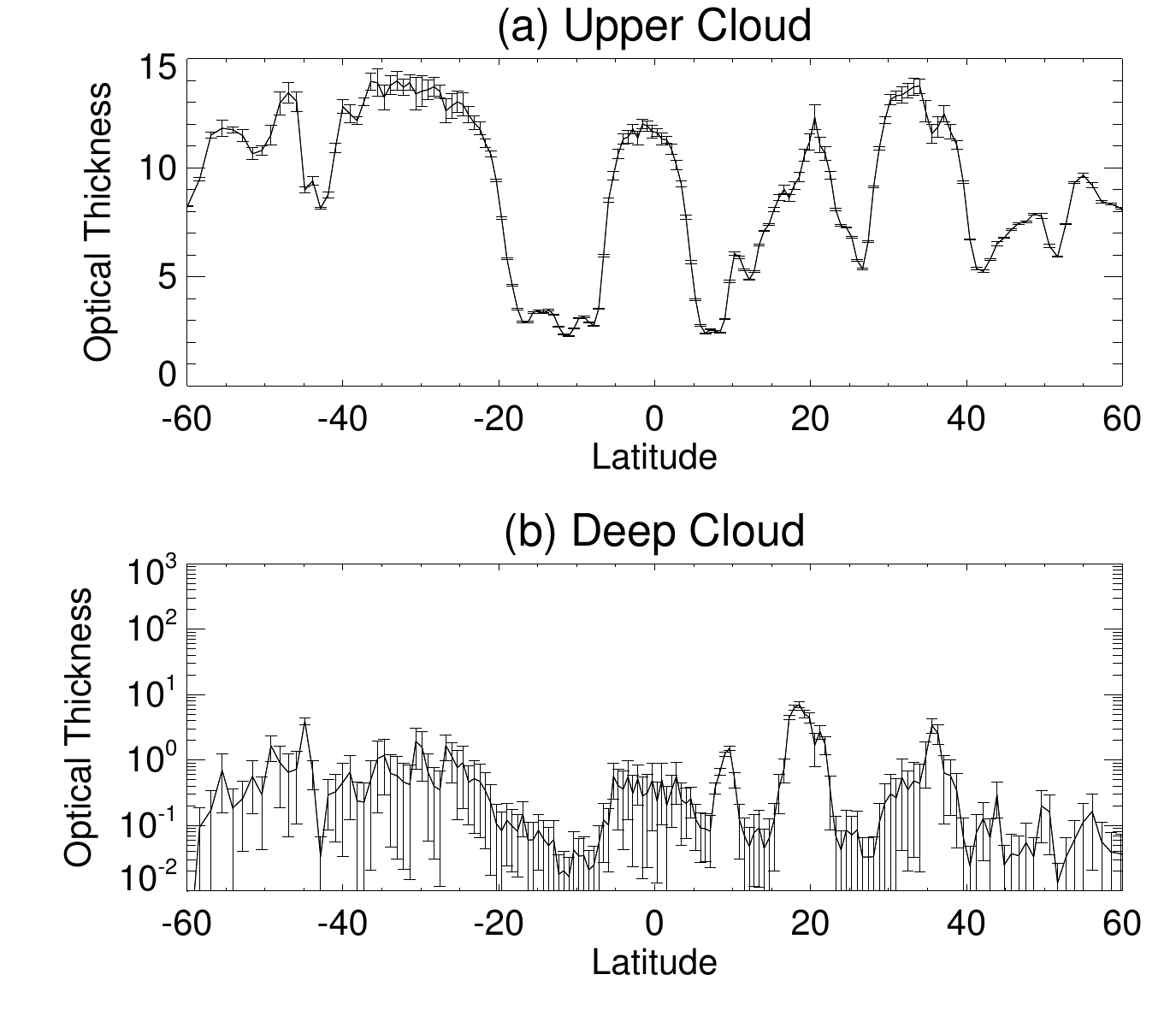}
\caption{Retrieved cloud optical thicknesses as a function of latitude, using $g=0.4$ instead of $g=0.8$ for the upper cloud.}
\label{fig:cloud_latitude2}
\end{figure}

Section~\ref{sec:asymmetry} showed that decreasing the asymmetry parameter, $g$, can act to significantly increase the retrieved gaseous abundances in the cloudy parts of the planet while having a negligible effect on the abundances in the cloud-free regions. This therefore has the potential to solve the problem posed in Section~\ref{sec:ch3danalysis} without having to invoke the presence of a deep cloud deck. With $g=0.8$, the retrieved CH\textsubscript{3}D abundance in the EqZ was 0.09 ppm, compared to 0.15 ppm in the SEB. If instead, we set $g=0.4$, the EqZ abundance increases to 0.16, while the SEB abundance remains 0.15 ppm, removing the discrepancy between the two regions. This is shown further by Figure~\ref{fig:cloud_latitude2}, which shows the same latitudinal cloud profiles as Figure~\ref{fig:cloud_latitude} for the case where $g=0.4$. Comparing these two figures shows that the change in scattering parameters significantly affects the latitudinal profile. The deep cloud is relatively transparent everywhere, with little difference between the belts and the zones. It is therefore clear that the conclusions about the existence of a cloud deck at 5 bar are fairly sensitive to the assumptions made about the scattering properties of the cloud deck at 0.8 bar. 

\subsection{Summary}

In this section, we have shown that the shape of the CH\textsubscript{3}D 5-\textmu m absorption features varies with latitude. We initially showed that this can be modelled in two different ways; either by allowing the CH\textsubscript{3}D abundance to vary spatially, or by allowing the opacity of a 5-bar tropospheric cloud to vary. We ruled out the former option on the grounds that CH\textsubscript{3}D should have a constant VMR throughout the troposphere, so we therefore accept the latter option. This suggests that there is an optically thick deep cloud layer present in the EqZ. We then performed a full latitudinal retrieval, where the opacity of this deep cloud was independently retrieved at each latitude, and we found that the deep cloud is opaque in the zones and transparent in the belts. 

It should be noted, however, that these conclusions are sensitive to the scattering properties of the upper cloud deck. The above conclusions are based on an asymmetry parameter of $g=0.8$, which was the best-fit value found in Section~\ref{sec:asymmetry}. However, if the asymmetry parameter is reduced to $g=0.4$, then the conclusions about the presence of a deep cloud are no longer valid. While $g=0.8$ did produce a better fit to the data than $g=0.4$, a value of 0.4 is still within a plausible range of values. We therefore conclude that while there is some evidence for the presence of a deep cloud, it is weakened by the degeneracies with the cloud scattering properties.

%%%%%%%%%%%%%%%%%%%%%%%%%%%%%%%%%%%%%%%%%%%%%
%%%%%%%%%%%%%%%%%%%%%%%%%%%%%%%%%%%%%%%%%%%%%

\section{GeH\textsubscript{4}}
\label{sec:geh4}

\subsection{Introduction}
\label{sec:geh4_introduction}

Germane (GeH\textsubscript{4}) is a molecular species that only exists in thermochemical equilibrium deep in Jupiter's atmosphere~\citep{taylor04}. In the 298-2000 K range, GeH\textsubscript{4} is modelled to be the second most abundant Ge species (after GeS), but as the altitude increases and the temperature decreases, it is destroyed in favour of either GeS or GeSe~\citep{fegley94}. Because of this, GeH\textsubscript{4} was not expected to be present in observable quantities in the upper troposphere at the pressure levels that are probed by the majority of the infrared spectrum. However, the 5-\textmu m atmospheric window probes much deeper into the atmosphere, leading~\citet{corice72} to predict that GeH\textsubscript{4} absorption features might be visible in this part of the spectrum.

GeH\textsubscript{4} has two infrared active fundamental bands, $\nu_3$ (4.737 \textmu m) and $\nu_4$ (12.210 \textmu m). Fortunately, $\nu_3$ is located in the middle of the 5-\textmu m window, providing the perfect opportunity to detect the disequilibrium species. It was the Q-branch of the $\nu_3$ fundamental that was first identified as a jovian GeH\textsubscript{4} feature by~\citet{fink78}, using observations from the Kuiper Airborne Observatory (KAO). They derived a tropospheric GeH\textsubscript{4} abundance of 0.6 ppb. The same spectral feature was subsequently observed in data from the Infrared Interferometer Spectrometer (IRIS) on the Voyager 1 spacecraft~\citep{kunde82,drossart82a}. Using new observations from KAO,~\citet{bjoraker86a} identified additional GeH\textsubscript{4} features, corresponding to the R3 and R6 transitions. In addition, they identified the Q-branch of the $\nu_1$ fundamental at 4.738 \textmu m; although this band is `forbidden' under symmetry considerations, weak absorption lines were identified in laboratory spectra by~\citet{lepage81}.

For these earlier analyses, the available spectroscopic data was restricted to a single isotopic species, \textsuperscript{74}GeH\textsubscript{4}. In fact, there are five naturally occurring isotopes of germanium: \textsuperscript{70}Ge, \textsuperscript{72}Ge, \textsuperscript{73}Ge, \textsuperscript{74}Ge and \textsuperscript{76}Ge. The terrestrial relative abundances are 20.6\%, 27.5\%, 7.8\%, 36.5\% and 7.7\% respectively~\citep{berglund11}.

\citet{bezard02} was the first study to include line data for all five isotopic species of GeH\textsubscript{4}. They used theoretical line data which they scaled to match observations for a single isotope, and assumed telluric relative abundances. Their derived VMR of 0.45 ppb was 35--50\% lower than previous estimates, which they attributed to the use of the additional isotopes. We take a similar approach to the GeH\textsubscript{4} line data as \citet{bezard02}. We generated theoretical line lists for each isotope using the Spherical Top Data System~\citep{wenger98}, and then scaled these to match the observed \textsuperscript{74}GeH\textsubscript{4} data from the GEISA database~\citep{jacquinet-husson99}. We also assume telluric relative abundances. As in previous studies, we assume that GeH\textsubscript{4} is well-mixed in the troposphere.

Previous studies of GeH\textsubscript{4} have generally been restricted to a single region of the planet: either an average over the centre of the disc~\citep{fink78,bjoraker86a} or a focus on the bright NEB~\citep{kunde82,bezard02}. The exception to this is~\citet{drossart82a}, who found that within a factor of 2, the same GeH\textsubscript{4} abundance was able to fit the range of spectra from 30$^{\circ}$S to 30$^{\circ}$N. We use the improved spectral resolution afforded by CRIRES to repeat this latitudinal study and to extend it to higher latitudes.

\subsection{Analysis}

\subsubsection{Identification of GeH\textsubscript{4} features}
\label{sec:geh4_features}

Our first task is to identify the GeH\textsubscript{4} absorption features in the 5-\textmu m spectra. Based on previous studies, we know that there is a strong Q-branch feature at 4.737 \textmu m. In a preliminary investigation, we determined that this absorption feature was particularly clear in the bright SEB; we therefore used this spatial region in order to search for additional GeH\textsubscript{4} absorption features. 

\begin{figure*}
\centering
\includegraphics[width=15cm]{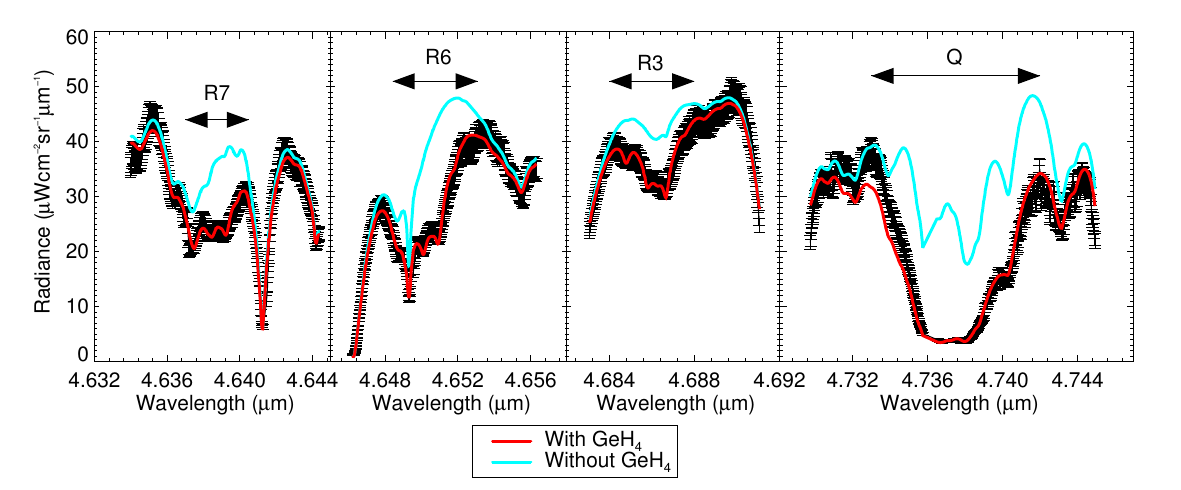}
\caption{Identification of GeH\textsubscript{4} absorption features in the 5-\textmu m window. CRIRES data is shown in black and is taken from the SEB. The best fits obtained from two retrievals are shown: with GeH\textsubscript{4} present (red) and with no GeH\textsubscript{4} present (blue).}
\label{fig:geh4_spectralfeatures}
\end{figure*}

The identifiable GeH\textsubscript{4} absorption features in the CRIRES spectra are shown in Figure~\ref{fig:geh4_spectralfeatures}. The NEMESIS retrieval algorithm was used to model the data, and the best-fit is shown in red. In order to highlight the shape and location of the GeH\textsubscript{4} absorption features, the blue line shows what the spectrum would look like in the absence of GeH\textsubscript{4}. Each feature is labelled, and is comprised of a blend of the five isotopologues. 

The Q, R3 and R6 bands have been identified in previous studies~\citep{fink78,bjoraker86a}. In addition, we identify the R7 line, located at 4.639 \textmu m. From each of these four spectral features, we retrieve the following GeH\textsubscript{4} abundances: 0.51 ppb (R7), 0.62 ppb (R6), 0.45 ppb (R3), 0.58 ppb (Q). These small differences could reflect genuine differences, due to each line probing a slightly different pressure level, or they could be due to inaccuracies in the line data. As each segment was observed at a different time, and therefore a different longitude, a simultaneous retrieval of all four features is not possible.

\subsubsection{Spatial variability}
\label{sec:geh4_spatial}

Having identified GeH\textsubscript{4} absorption features in one discrete region of the planet (the SEB), we now consider how this varies with latitude. We will focus on the most prominent GeH\textsubscript{4} feature, the Q-branch at 4.734-4.742 \textmu m.

\begin{figure*}
\centering
\includegraphics[width=17cm]{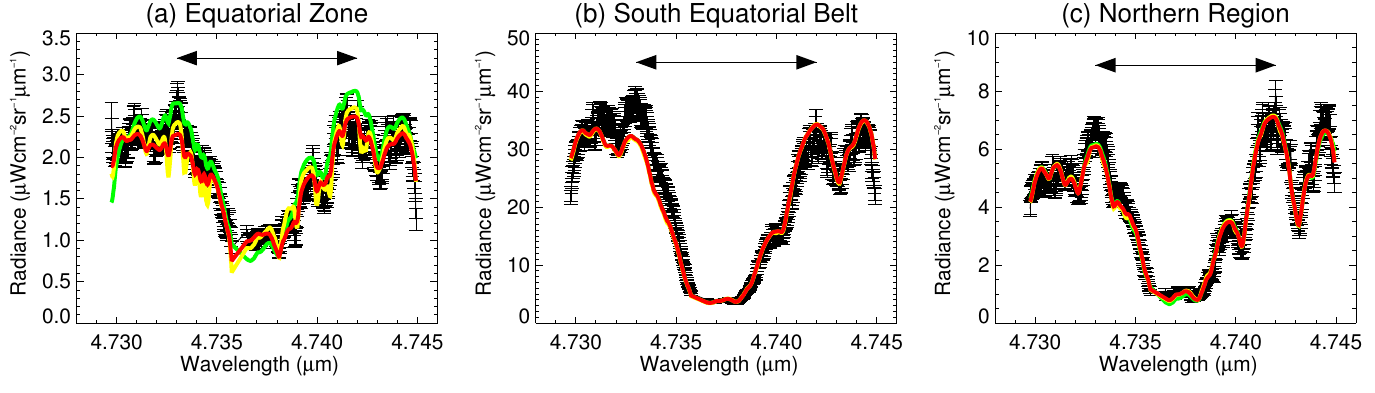}
\caption{GeH\textsubscript{4} absorption feature fits in the Equatorial Zone and South Equatorial Belt. The CRIRES oservations are shown in black. The coloured lines show the best fit that can be achieved using different assumptions. Red: single cloud deck with $g=0.8$, GeH\textsubscript{4} allowed to vary. Yellow: single cloud deck with $g=0.4$, GeH\textsubscript{4} allowed to vary. Green: upper cloud deck with $g=0.8$, additional deep cloud deck, GeH\textsubscript{4} held fixed.} 
\label{fig:geh4_fits}
\end{figure*}

Figure~\ref{fig:geh4_fits} shows the same absorption feature for three different regions of the planet: the EqZ, the SEB and a ``Northern Region'', corresponding to 50-55$^{\circ}$N. Comparing the data, shown in black, from the three regions highlights some differences: the absorption feature appears to be deepest in the SEB and shallowest in the EqZ. This would appear to suggest that there is less GeH\textsubscript{4} present in the EqZ. However, the previous sections have already demonstrated a degeneracy between cloud properties and retrieved gaseous abundances. Section~\ref{sec:asymmetry} showed that the depth of an absorption feature can depend on the scattering properties of the main cloud deck, and Section~\ref{sec:ch3d} showed that it can also depend on the presence/absence of a deep cloud deck. Since the primary difference in line shape is between the EqZ and SEB, which we expect to have very different cloud structures, it is likely that clouds play an important role. 

In order to explore whether or not the CRIRES observations provide evidence for spatial variability in GeH\textsubscript{4}, we performed a series of retrievals on these three spectra, making different assumptions about the cloud structure. The red line shows the case where there is a single upper cloud deck, with an asymmetry parameter $g=0.8$. The retrieved GeH\textsubscript{4} abundances are 0.19 ppb (EqZ), 0.58 ppb (SEB) and 0.44 ppb (Northern Region); in this case, the shallow feature in the EqZ does indeed lead to a lower retrieved abundance.

The yellow line in Figure~\ref{fig:geh4_fits} shows the case where the asymmetry parameter of the cloud is changed to 0.4. Because the SEB and Northern Region have low optical thicknesses, this has a negligible effect of those retrievals (now 0.59 ppb and 0.47 ppb respectively). There is a significant effect for the EqZ, increasing the retrieved abundance from 0.19 ppb to 0.47 ppb. Changing the asymmetry parameter (i.e. making the cloud particles less forward scattering) can therefore account for much of the apparent difference between the EqZ and the SEB.

Alternatively, the difference can be accounted for by the presence/absence of a deep cloud. In Section~\ref{sec:ch3d}, we fixed the CH\textsubscript{3}D abundance to the SEB value, and we were able to fit the EqZ by adding a deeper cloud deck. We repeat the same process here: upper cloud is returned to an asymmetry parameter of 0.8, the GeH\textsubscript{4} abundance is fixed to the SEB value (0.58 ppb) and the opacity of a deep cloud at 5 bar is allowed to vary in the retrievals. The green line in Figure~\ref{fig:geh4_fits} shows the fits obtained for this case. The retrieved deep cloud is opaque in the EqZ and transparent in the SEB, just as in Section~\ref{sec:ch3d}. This shows that the deep cloud structure studied in Section~\ref{sec:ch3d} can account for the difference in line shape between the EqZ and SEB.

\subsubsection{Latitudinal retrieval}
\label{sec:geh4_latitude}

\begin{figure}
\centering
\includegraphics[width=9cm]{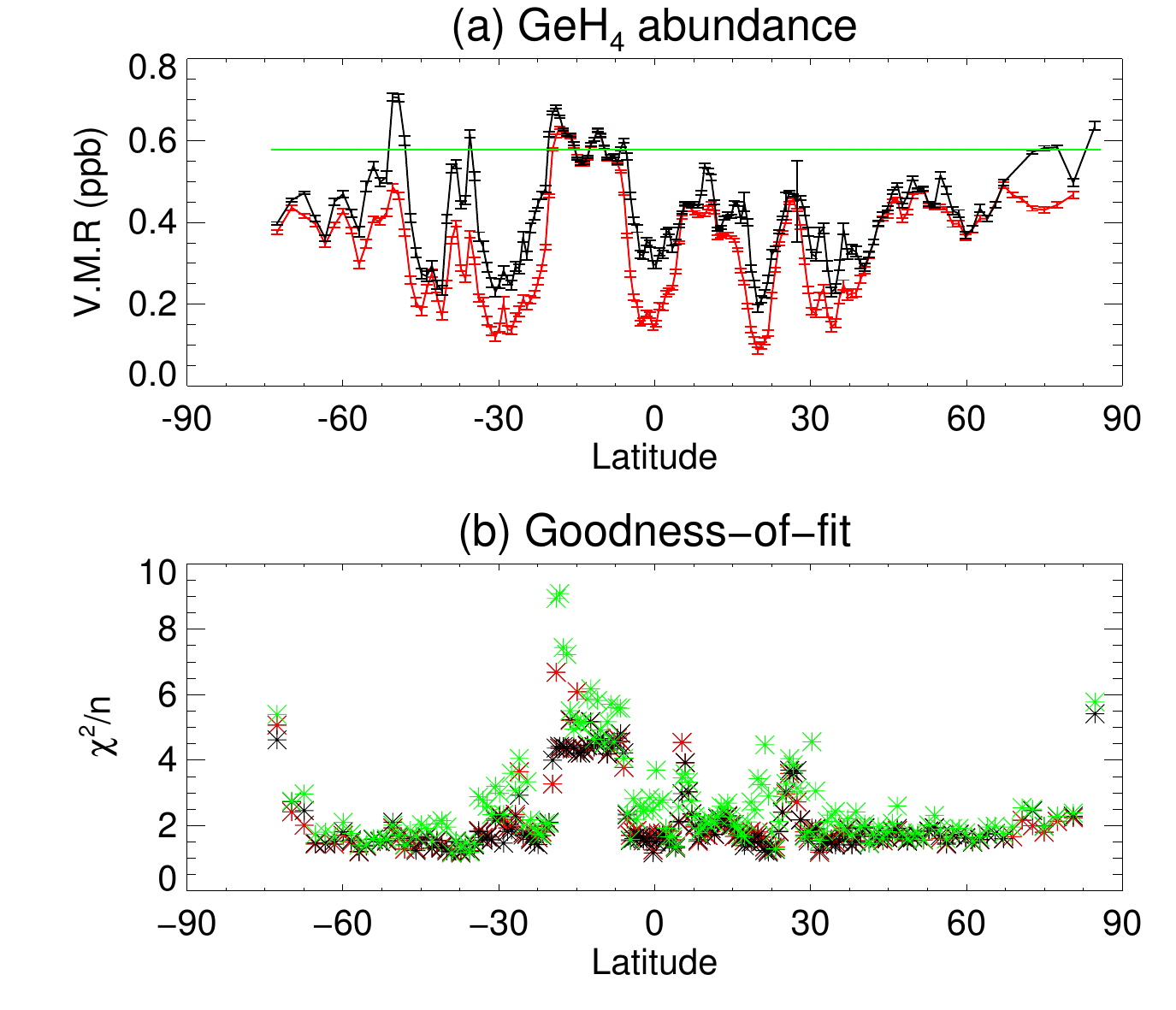}
\caption{GeH\textsubscript{4} abundance and goodness-of-fit values as a function of latitude, for three different assumptions. Red: single cloud deck with $g=0.8$, GeH\textsubscript{4} allowed to vary. Black: additional deep cloud deck with opacities derived from Section~\ref{sec:ch3d}, GeH\textsubscript{4} allowed to vary. Green: deep cloud deck allowed to vary, GeH\textsubscript{4} held fixed at 0.58 ppb.}
\label{fig:geh4_latitude}
\end{figure}

Section~\ref{sec:geh4_spatial} showed that there is no evidence for spatial variability in GeH\textsubscript{4} between three discrete regions of the planet, as any differences in line shape can be accounted for by changes in the cloud structure alone. In this section, we extend the analysis to cover the full range of latitudes available. Figure~\ref{fig:geh4_latitude} shows the GeH\textsubscript{4} abundance as a function of latitude, retrieved with different assumptions. Alongside this is the goodness-of-fit (\textchi\textsuperscript{2}/n), which describes the quality of the fit at each latitude point.  As with Figure~\ref{fig:geh4_fits}, the three colours correspond to different assumptions made in the retrievals.

The red line corresponds to the simplest case, where there is a single upper cloud deck (with $g=0.8$) and the GeH\textsubscript{4} abundance is allowed to vary, i.e. the same as the red line in Figure~\ref{fig:geh4_fits}. Section~\ref{sec:geh4_spatial} showed that this simple case led to a variability in the retrieved GeH\textsubscript{4} abundance, with a depletion in the EqZ and an enhancement in the SEB. Figure~\ref{fig:geh4_latitude} shows that this applies at all latitudes, and so this simple assumption leads to an apparent belt-zone variability.

The black line corresponds to the case where an additional deep cloud is included in the model and its latitudinal profile is fixed to the values retrieved in Section~\ref{sec:ch3d} (shown in Figure~\ref{fig:cloud_latitude}). It should be noted that this previously determined deep cloud latitudinal profile relates to a different longitude, so it is not necessarily expected to be identical to the cloud profile at this longitude. Nevertheless, the overall latitudinal trend is likely to be very similar and it provides an insight into the role of the deep cloud on the retrieved abundances. Once again, the GeH\textsubscript{4} abundance is allowed to vary in the retrievals. It can be seen in Figure~\ref{fig:geh4_latitude}a that the presence of this deep cloud increases the retrieved abundances in the zones, decreasing the apparent belt-zone variability shown by the red line. Figure~\ref{fig:geh4_latitude}b shows that this change does not alter the quality of the fits.

For the green line, the GeH\textsubscript{4} abundance is held fixed as a function of latitude and the opacity of the deep cloud is allowed to vary, instead of being held fixed at the values determined in Section~\ref{sec:ch3d}. This is the same as the green lines in Figure~\ref{fig:geh4_fits}. As in Section~\ref{sec:geh4_spatial}, the GeH\textsubscript{4} abundance chosen is the best-fit value from the SEB (0.58 ppb). Since the SEB is relatively cloud-free, there are no complications from the cloud structure and this should represent the `true' abundance. The goodness-of-fit in Figure~\ref{fig:geh4_latitude} shows that this constant abundance is able to provide a similar fit to the data at all latitudes. Section~\ref{sec:geh4_spatial} previously showed that variability in the deep cloud opacity could account for the difference in line shape between three discrete regions of the planet, and Figure~\ref{fig:geh4_latitude} shows that this remains true across the full range of latitudes.

\subsection{Summary}
\label{sec:geh4_conclusions}

In this section, we have identified four GeH\textsubscript{4} absorption features in the CRIRES observations, one of which has not been previously identified. Using the strongest absorption feature, we searched for latitudinal variability and found that there are some differences in the line depth between the EqZ, the SEB and the Northern Region of the planet. However, this difference in line shape does not necessarily indicate a difference in the abundance of GeH\textsubscript{4}. We found that the variation in line shape could be accounted for by Jupiter's tropospheric cloud structure. This analysis was then extended to all latitudes, and it was found that a single fixed GeH\textsubscript{4} abundance could be used to fit all spectra without compromising the quality of the fits. The CRIRES observations therefore do not provide any evidence for spatial variability in GeH\textsubscript{4}. This highlights the high level of degeneracy present in the problem. The deep cloud structure, which has not been included in an any previous studies of GeH\textsubscript{4}, significantly complicates the analysis. 

%%%%%%%%%%%%%%%%%%%%%%%%%%%%%%%%%%%%%%%%%%%%%
%%%%%%%%%%%%%%%%%%%%%%%%%%%%%%%%%%%%%%%%%%%%%

\section{AsH\textsubscript{3}}
\label{sec:ash3}

\subsection{Introduction}

Arsine (AsH\textsubscript{3}) is the most abundant arsenic gas on Jupiter. Deep in the atmosphere it exists in thermochemical equilibrium but, in the absence of vertical motion, it precipitates into As\textsubscript{4}(s) or As\textsubscript{2}S\textsubscript{2}(s) at $\sim$400 K ~\citep{fegley94}. As with GeH\textsubscript{4}, this means that the 5-\textmu m atmospheric window is an ideal spectral region to search for the species.

AsH\textsubscript{3} has two fundamental bands within the 5-\textmu m window: $\nu_1$ at 4.728 \textmu m and $\nu_3$ at 4.704 \textmu m. Of these, the $\nu_3$ band is stronger and its Q-branch led to the first detection of AsH\textsubscript{3} on Jupiter by~\citet{noll89}, using spectra obtained from the United Kingdom Infrared Telescope. They initially estimated the deep molar abundance to be $0.7\substack{+0.7 \\ -0.4}$ ppb, which was then revised down to $0.22\pm0.11$ ppb the following year, after acquisition of new airborne and ground-based spectra and additional laboratory line data~\citep{noll90}.  More recently, \citet{bezard02} used data from the Canada-France-Hawaii Telescope to obtain an abundance of 0.24 ppb. The Q-branch of the $\nu_3$ band remains the only AsH\textsubscript{3} feature to have been identified on Jupiter; we will search for evidence of additional AsH\textsubscript{3} absorption lines. As with GeH\textsubscript{4}, we assume that AsH\textsubscript{3} is well-mixed in the troposphere.

Out of the previous studies of AsH\textsubscript{3}, only~\citet{noll90} studied more than one region of the planet. They made observations of both the EqZ and the NEB and found that these spectra matched to within the noise levels. As the CRIRES observations provide both high spectral resolution and high spatial resolution, we will extend this analysis to cover all latitudes.

\subsection{Analysis}

\subsubsection{Identification of AsH\textsubscript{3} features}

As with GeH\textsubscript{4}, we first seek to simply identify the AsH\textsubscript{3} absorption features in the 5-\textmu m spectrum. The only previously identified feature is the Q-branch at 4.704 \textmu m; in a preliminary investigation we determined that this feature was particularly strong at high latitudes (50-55$^{\circ}$N) so we use this `Northern Region' in order to search for additional absorption lines.

\begin{figure*}
\centering
\includegraphics[width=12cm]{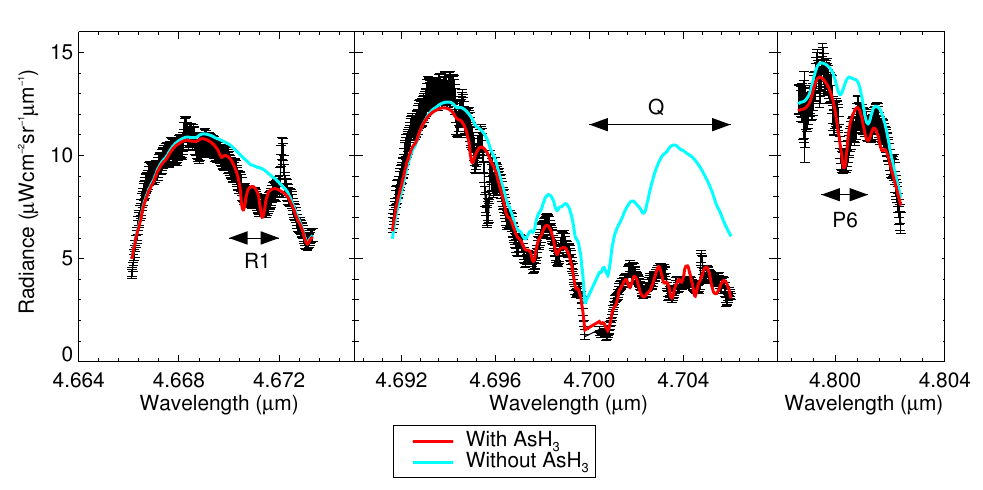}
\caption{Identification of AsH\textsubscript{3} absorption features in the 5-\textmu m window. CRIRES data is shown in black, and is taken from the Northern Region of the planet. The best fits obtained from two retrievals are shown: with AsH\textsubscript{3} present (red) and with no AsH\textsubscript{3} present (blue).}
\label{fig:ash3_spectralfeatures}
\end{figure*}

We were able to identify two additional absorption features, corresponding to the R1 and P6 lines of the $\nu_3$ band~\citep{dana93}. These lines are shown in Figure~\ref{fig:ash3_spectralfeatures}, alongside the Q-branch. The CRIRES data is shown in black. The blue lines show the best fit that can be achieved when no AsH\textsubscript{3} is present in the atmosphere and the red lines show the best fit that can be achieved when the AsH\textsubscript{3} abundance is allowed to vary. As in Section~\ref{sec:geh4_features}, the cloud opacity of the single cloud deck and the gaseous abundances were also allowed to vary. In each case, the addition of AsH\textsubscript{3} significantly improves the fit. We obtain the following abundances from each retrieval: 0.60 ppb (R1), 0.49 ppb (Q) and 0.46 ppb (P6).

\subsubsection{Spatial variability}
\label{sec:spatial_ash3}

\begin{figure*}
\centering
\includegraphics[width=17cm]{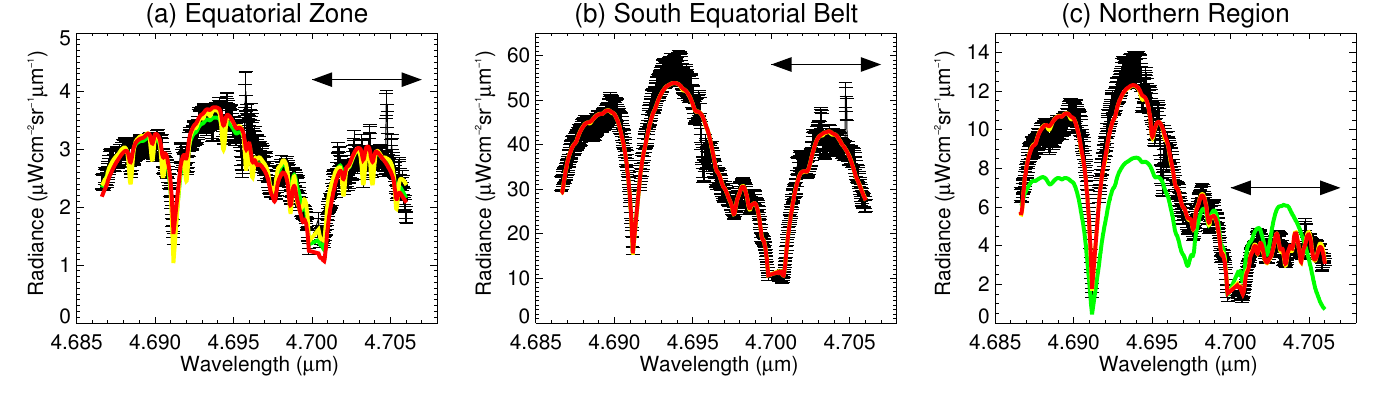}
\caption{AsH\textsubscript{3} absorption feature fits at different latitudes. The CRIRES observations are shown in black. The coloured lines show the best fit that can be achieved using different assumptions. Red: single cloud deck with $g=0.8$, AsH\textsubscript{3} allowed to vary. Yellow: single cloud deck with $g=0.4$, AsH\textsubscript{3} allowed to vary. Green: upper cloud deck with $g=0.8$, additional deep cloud deck, AsH\textsubscript{3} held fixed.}
\label{fig:ash3_fits}
\end{figure*}

We now consider how the AsH\textsubscript{3} lineshape varies with latitude, and how this affects the retrieved abundances. As in previous studies, we focus on the most prominent AsH\textsubscript{3} absorption feature in the 5-\textmu m window, the Q-branch of the $\nu_3$ fundamental, located at 4.704 \textmu m. Figure~\ref{fig:ash3_fits} shows this spectral region for the three regions of the planet considered in Section~\ref{sec:geh4}. Unlike the previous GeH\textsubscript{4} section, the EqZ and the SEB have similar spectral shapes. In this case, however, there is significant variation between these spectra and the spectrum from the Northern Region. The absorption feature at 4.704 \textmu m is considerably flatter in Northern Region than it is in the SEB and the EqZ, suggesting a higher abundance of AsH\textsubscript{3} at high latitudes. 

This is confirmed by a simple NEMESIS retrieval, shown in red in Figure~\ref{fig:ash3_fits}. This fit is analogous to the red line in Figure~\ref{fig:geh4_fits}, where the cloud structure consists of a single upper cloud with $g=0.8$. With these assumptions, we obtain the following retrieved abundances for AsH\textsubscript{3}: 0.01 ppb in the EqZ, 0.02 ppb in the SEB and 0.49 ppb in the Northern Region. In the case of the EqZ and the SEB, the retrieved abundances are so low that they are essentially zero.

In Section~\ref{sec:geh4}, we found that the apparent change in GeH\textsubscript{4} lineshape could be accounted for by variations in the cloud structure, rather than changes in the GeH\textsubscript{4} abundance itself. However, clouds are unlikely to be responsible here, since the EqZ and the SEB have very different cloud structures, and yet the AsH\textsubscript{3} features have similar spectral shapes. The observed effect is also unlikely to be due to limb darkening at high altitudes. Firstly, previous studies have shown that Jupiter's tropospheric clouds are highly scattering, which minimises the effect of limb darkening~\citep{roos-serote06,giles15}. Secondly, even if there was a significant limb darkening effect, we would to see the opposite phenomenon. As we move to higher latitudes, the emission angle increases, and we therefore probe higher in the atmosphere, where (as a disequilibrium species) AsH\textsubscript{3} is \textit{less} abundant, not more.

Nevertheless, in order to fully explore any degeneracy, we repeated the different retrievals performed in Section~\ref{sec:geh4}. The yellow line in Figure~\ref{fig:ash3_fits} shows the fits that are obtained when the asymmetry parameter of the upper cloud is changed from 0.8 to 0.4. The retrieved abundance in the EqZ is increased from 0.01 ppb to 0.05 ppb, and the SEB and Northern abundances remain essentially unchanged at 0.02 ppb and 0.50 ppb respectively. We are unable to close the gap between the equatorial regions and the high latitudes by changing the scattering properties of the upper cloud.

We then consider the effect of a deep cloud. The AsH\textsubscript{3} were held constant at the SEB level (0.02 ppb) and the opacity of a deep cloud was allowed to vary. The results are shown by the green line in Figure~\ref{fig:ash3_fits}. As in Section~\ref{sec:geh4}, the EqZ and SEB can be fit using the same AsH\textsubscript{3} abundance, provided that the deep cloud is optically thick in the EqZ. However, it is clear from Figure~\ref{fig:ash3_fits}(c) that this constant abundance cannot be applied to the Northern Region of the planet. We therefore conclude that the spatial variation in the AsH\textsubscript{3} lineshape is indicative of genuine latitudinal variability in the abundance of AsH\textsubscript{3}.

\subsubsection{Latitudinal retrieval}

\begin{figure}
\centering
\includegraphics[width=9cm]{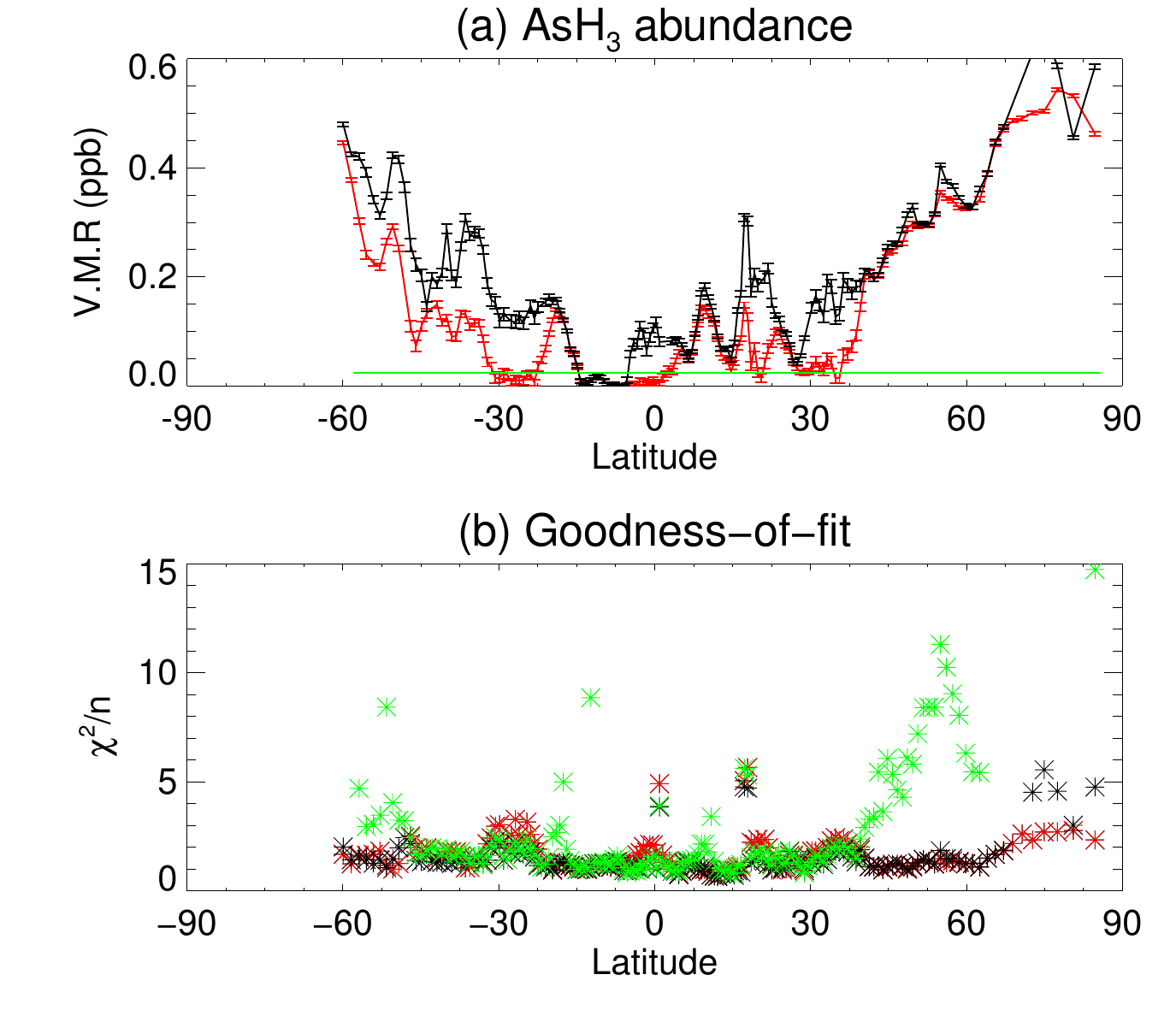}
\caption{AsH\textsubscript{3} abundance and goodness-of-fit values as a function of latitude, for three different assumptions. Red: single cloud deck with $g=0.8$, AsH\textsubscript{3} allowed to vary. Black: additional deep cloud deck with opacities derived from Section~\ref{sec:ch3d}, AsH\textsubscript{3} allowed to vary. Green: deep cloud deck allowed to vary, AsH\textsubscript{3} held fixed at 0.02 ppb.}
\label{fig:ash3_latitude}
\end{figure}

Having established that there is evidence for spatial variability in AsH\textsubscript{3}, we now extend the analysis from three discrete sections to a full latitudinal retrieval of AsH\textsubscript{3}. Figure~\ref{fig:ash3_latitude} shows the latitudinal distribution of AsH\textsubscript{3} alongside the goodness-of-fit values. The three different colours correspond to the same assumptions described in Section~\ref{sec:geh4_latitude}.

For the latitudinal distribution shown by the red line, there is a single upper cloud deck with $g=0.8$ and the AsH\textsubscript{3} abundance is allowed to vary, i.e. the same as the red line in Figure~\ref{fig:ash3_fits}. Section~\ref{sec:spatial_ash3} showed that this simple case produced similar retrieved abundances in the EqZ and SEB and an enhanced abundance in the Northern Region. This pattern is confirmed by the full latitudinal retrieval; there is a roughly symmetrical distribution, with an increased abundance at high latitudes. 

The black line shows the case where an additional deep cloud is included in the model, with the latitudinal profile obtained from Section~\ref{sec:ch3d}. While the precise values of the retrieved abundances do change slightly with the inclusion of this additional cloud deck, the overall trend of an enhancement an high latitudes remains true. The errors shown in Figure~\ref{fig:ash3_latitude} represent the formal errors on the retrieval; the difference between the red and black lines shows that these formal errors are significantly smaller than the errors due to varying assumptions about cloud structure. As with Section~\ref{sec:geh4_latitude} for GeH\textsubscript{4}, the inclusion of an additional cloud does not alter the quality of the fits. 

This is in contrast to the final case, shown by the green line. In this case, the AsH\textsubscript{3} abundance is held fixed at the SEB value of 0.02 ppb and the deep cloud is allowed to vary. Section~\ref{sec:spatial_ash3} showed that this can provide a good fit in the EqZ and SEB, but fails to provide a good fit for the Northern Region. Figure~\ref{fig:ash3_latitude} extends this to all latitudes and shows that this constant abundance can provide a good fit between 40$^{\circ}$S and 40$^{\circ}$N, i.e. there is no evidence for belt-zone variability. However, the fit worsens considerably at high latitudes; an abundance of 0.02 ppb simply cannot fit these regions, regardless of the opacity of the deep cloud. This leads to the conclusion that there is a genuine enhancement at high latitudes. 

\begin{figure}
\centering
\includegraphics[width=9cm]{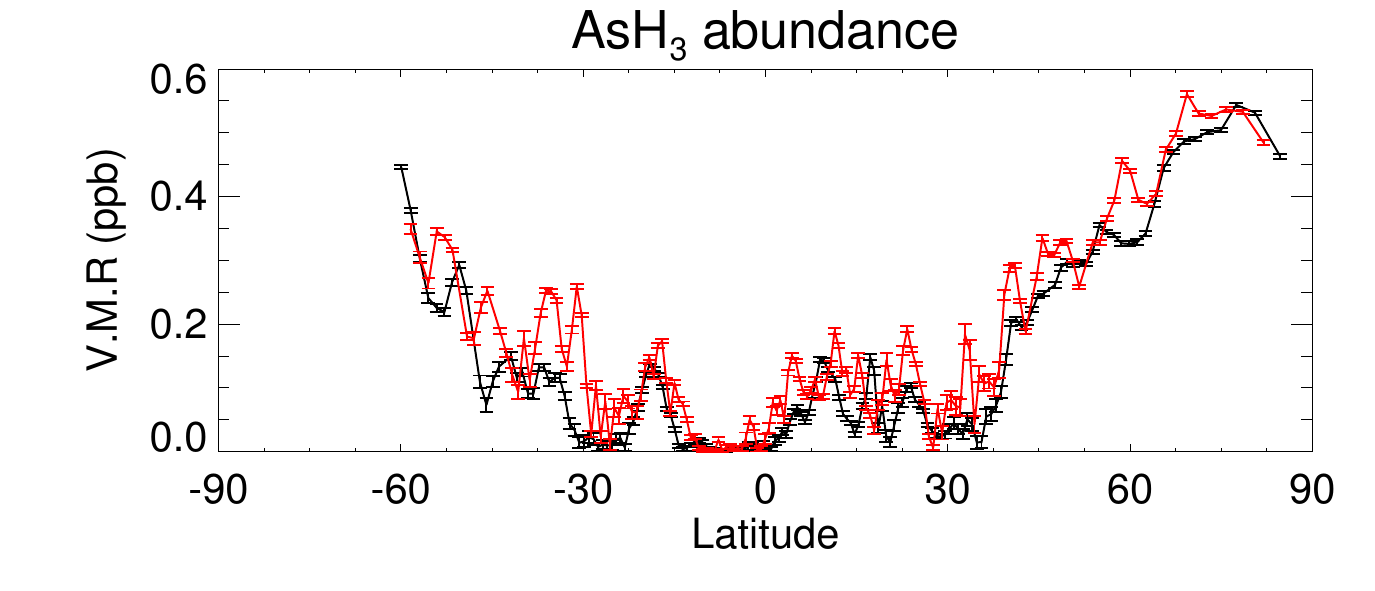}
\caption{AsH\textsubscript{3} abundance as a function of latitude from the 12 November 2012 dataset (black) and the 1 January 2013 dataset (red). In both cases, a single cloud deck with $g=0.8$ is assumed and the AsH\textsubscript{3} abundance is allowed to vary. The two datasets produce consistent results.}
\label{fig:ash3_latitude_nov_jan}
\end{figure}

Figure~\ref{fig:ash3_latitude} is restricted to the data from 12 November 2012. However, Figure~\ref{fig:ash3_latitude_nov_jan} shows that the 1 January 2013 dataset produces consistent results, and that the overall conclusion of an AsH\textsubscript{3} enhancement at high latitudes is reproducible.

\subsection{Summary}

In this section, we have identified three AsH\textsubscript{3} absorption features in the CRIRES observations, two of which have not been previously identified. Using the strongest absorption feature, we searched for latitudinal variability and found that there are significant differences in the line shape between spectra from the equatorial regions (EqZ and SEB) and spectra from the far northern latitudes. These differences cannot be accounted for by Jupiter's cloud structure, and we therefore conclude that there is evidence for latitudinal variability in the abundance of AsH\textsubscript{3}. This was further shown by a latitudinal retrieval, which demonstrated that a constant AsH\textsubscript{3} abundance is inconsistent with the observations. Instead, the retrieved AsH\textsubscript{3} abundances have a roughly symmetrical latitudinal profile, with a minimum in the equatorial regions and enhanced abundances at high latitudes.

%%%%%%%%%%%%%%%%%%%%%%%%%%%%%%%%%%%%%%%%%%%%%
%%%%%%%%%%%%%%%%%%%%%%%%%%%%%%%%%%%%%%%%%%%%%

\section{PH\textsubscript{3}}
\label{sec:ph3}

\subsection{Introduction}

Like GeH\textsubscript{4} and AsH\textsubscript{3}, phosphine (PH\textsubscript{3}) is a disequilibrium species in Jupiter's troposphere. Deep in the atmosphere it is the primary phosphorus gas but at higher altitudes thermal decomposition produces PH, which then reacts with OH radicals to form P\textsubscript{4}O\textsubscript{6}~\citep{fegley94}. However, the solar abundance of P is much greater than the solar abundances of Ge or As~\citep{grevesse07}. This means that even though the VMR of PH\textsubscript{3} decreases rapidly with altitude, it is still present in observable quantities in the upper troposphere and is therefore detectable outside the 5-\textmu m window. 

Strong PH\textsubscript{3} absorption features can be seen at $\sim$ 5 \textmu m and at $\sim$10 \textmu m. PH\textsubscript{3} was simultaneously detected in both of these spectral regions. \citet{ridgway76} noted that including a solar abundance of PH\textsubscript{3} in their model improved their fit to 10-\textmu m observations of Jupiter from the McMath Solar Observatory. This was then confirmed by \citet{larson77}, who also found that a roughly solar abundance of PH\textsubscript{3} was able to fit their 5-\textmu m airborne observations.

The 5-\textmu m and 10-\textmu m spectral regions probe different pressure levels in Jupiter's troposphere. At 5-\textmu m, the spectrum is primarily sensitive to the lower troposphere (p\textgreater1 bar) whereas the 10-\textmu m observations are mostly sensitive to the upper troposphere, at pressures p\textless1 bar. This pressure difference allows us to start to constrain the vertical profile of PH\textsubscript{3}. \citet{fletcher09} modelled Jupiter's PH\textsubscript{3} abundance using a constant abundance of 1.86 ppm up to 1 bar, above which it drops off with a constant fractional scale height of 0.3. In this section, we will vary this profile via a single scaling parameter; because the CRIRES data is from the 5-\textmu m region, we are probing the deep abundance. 

Several previous studies have also considered the possibility of spatial variability in the deep PH\textsubscript{3} abundance, as observed in the 5-\textmu m window.~\citet{drossart82a} and~\citet{lellouch89} searched for a correlation between cloud opacity and PH\textsubscript{3} abundance, but did not find any evidence. More recently, ~\citet{drossart90} and~\citet{giles15} found evidence for a PH\textsubscript{3} enhancement at high latitudes compared to equatorial- and mid-latitudes. The spatially resolved 10-\textmu m studies show a different pattern; the retrieved fractional scale height shows a global maximum over the equator and global minima at the NEB and SEB~\citep{irwin04,fletcher09}.

\subsection{Analysis}

\subsubsection{Spatial variability}
\label{sec:spatial_ph3}

For GeH\textsubscript{4} and AsH\textsubscript{3}, the choice of spectral region was clear, as there were very few absorption features in the 5-\textmu m window. In contrast, PH\textsubscript{3} is one of the primary absorbers at 5-\textmu m, and there are many absorption features. In particular, the short-wavelength part of the 5-\textmu m spectrum (4.50-5.00 \textmu m) is dominated by PH\textsubscript{3} absorption. This might sound ideal, but in fact this prevents the accurate retrieval of PH\textsubscript{3}, as it becomes difficult to distinguish between PH\textsubscript{3} and other broad features such as the cloud opacity and the water vapour abundance. Instead, we look towards the long-wavelength edge, where PH\textsubscript{3} absorption contributes less to the continuum. We find that there is a clear, well-separated absorption feature at 5.071 \textmu m and we will use this as the basis of our study of the deep PH\textsubscript{3} abundance in the troposphere. 

\begin{figure*}
\centering
\includegraphics[width=17cm]{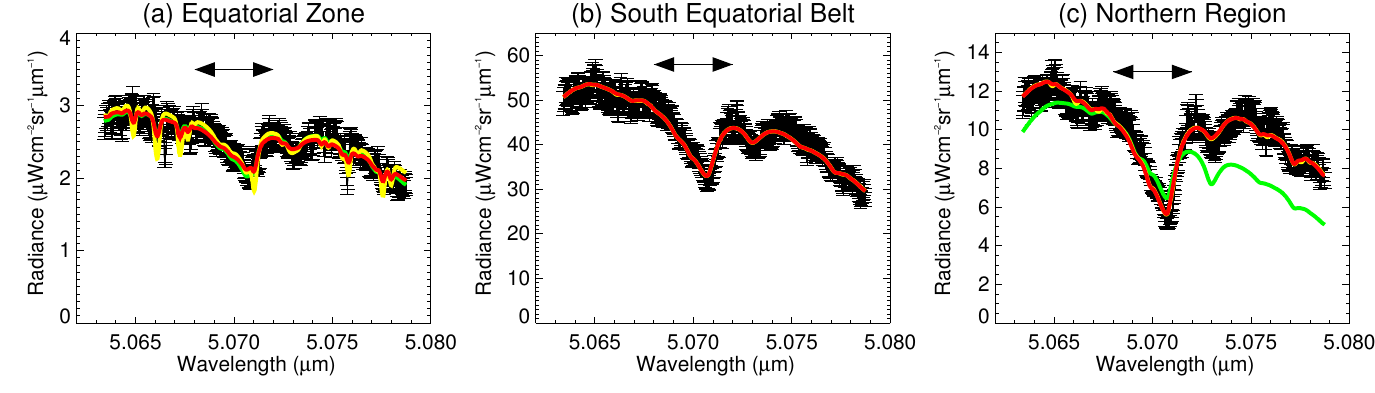}
\caption{PH\textsubscript{3} feature fits at different latitudes. The CRIRES observations are shown in black. The coloured lines show the best fit that can be achieved using different assumptions. Red: single cloud deck with $g=0.8$, PH\textsubscript{3} allowed to vary. Yellow: single cloud deck with $g=0.4$, PH\textsubscript{3} allowed to vary. Green: upper cloud deck with $g=0.8$, additional deep cloud deck, PH\textsubscript{3} held fixed.}
\label{fig:ph3_fits}
\end{figure*}

Figure~\ref{fig:ph3_fits} shows this absorption feature for the same three spatial regions used in Sections~\ref{sec:geh4} and \ref{sec:ash3}. The CRIRES spectra in Figure~\ref{fig:ph3_fits} show variation in the depth of the PH\textsubscript{3} absorption feature; the feature is considerably deeper in the Northern Region than in the EqZ, while the SEB spectrum is between these two extremes. 

The coloured lines show the best fit obtained for each spectrum, with the same assumptions as in Sections~\ref{sec:geh4} and~\ref{sec:ash3}: the red line shows the case where there is a single upper cloud with $g=0.8$ and the yellow line shows the case where there is a single upper cloud with $g=0.4$. In the first case (red), the retrieved abundances in the EqZ, SEB and Northern Region are 0.49 ppm, 0.76 ppm and 1.22 ppm respectively. In the second case (yellow), these become 0.81 ppm, 0.75 ppm and 1.21 ppm. As in Section~\ref{sec:ash3}, altering the scattering properties of the cloud can account for the discrepancy between the EqZ and the SEB but not between the equatorial regions and the polar regions. 

The green line in Figure~\ref{fig:ph3_fits} shows the case where the upper cloud has $g=0.8$, the PH\textsubscript{3} abundance is fixed to the SEB value and the opacity of a second deep cloud is allowed to vary. Again, the results are the same as for AsH\textsubscript{3} in Section~\ref{sec:ash3}. The presence of an optically thick deep cloud in the EqZ can account for the change in spectral shape between the EqZ and the SEB, but the same PH\textsubscript{3} abundance cannot be used to fit the regions at high latitude. We conclude that there is evidence for latitudinal variability in PH\textsubscript{3}.

\subsubsection{Latitudinal retrieval}

\begin{figure}
\centering
\includegraphics[width=9.cm]{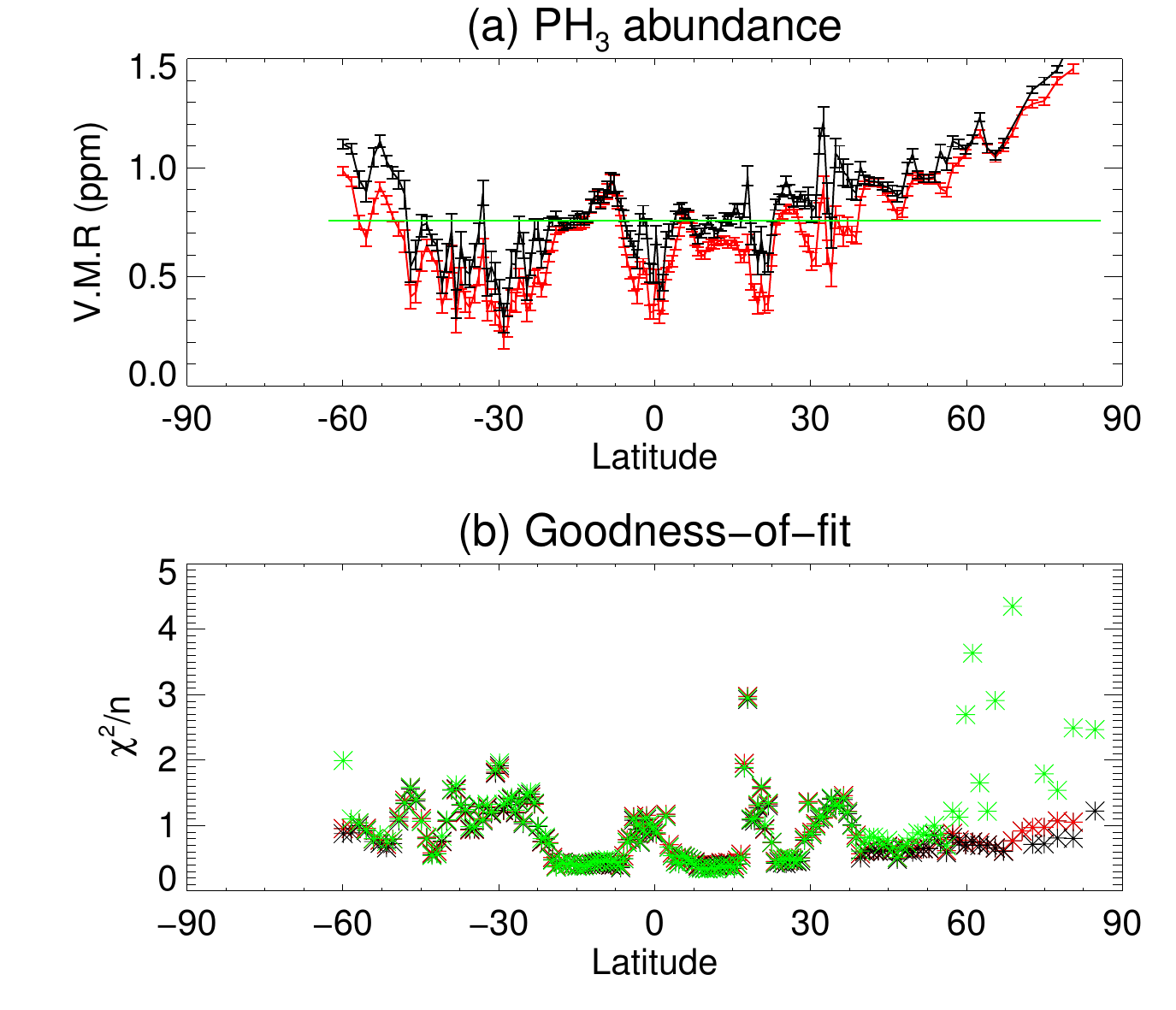}
\caption{PH\textsubscript{3} abundance and goodness-of-fit values as a function of latitude, for three different assumptions. Red: single cloud deck with $g=0.8$, PH\textsubscript{3} allowed to vary. Black: additional deep cloud deck with opacities derived from Section~\ref{sec:ch3d}, PH\textsubscript{3} allowed to vary. Green: deep cloud deck allowed to vary, PH\textsubscript{3} held fixed at 0.76 ppm.}
\label{fig:ph3_latitude}
\end{figure}

Having previously focussed on three discrete sections of the planet, we now turn to a full latitudinal retrieval of PH\textsubscript{3}. he results of these retrievals are shown in Figure~\ref{fig:ph3_latitude}; these results take the same form as Figure~\ref{fig:geh4_latitude} (for GeH\textsubscript{4}) and Figure~\ref{fig:ash3_latitude} (for AsH\textsubscript{3}) and the same conclusions can be drawn for PH\textsubscript{3} as for AsH\textsubscript{3}. 

As with AsH\textsubscript{3}, the black and red lines in Figure ~\ref{fig:ph3_latitude} both show a polar enhancement in PH\textsubscript{3}, whether or not the additional deep cloud from Section~\ref{sec:ch3d} is included in the model. The case where the PH\textsubscript{3} abundance is held fixed at 0.76 ppm is shown in green, and this also follows the same pattern as AsH\textsubscript{3}; at low latitudes, this model provides a good fit (i.e. there is no evidence for belt-zone variability), but the fit starts to worsen at high latitudes (in this case, above $\sim$60$^{\circ}$N). This agrees with the conclusion made in Section~\ref{sec:spatial_ph3} that there in evidence for an enhancement in PH\textsubscript{3} at high latitudes, with the  retrieved abundance varying from $\sim$0.5 ppm near the equator to 1.5 ppm at 80$^{\circ}$N. 

\begin{figure}
\centering
\includegraphics[width=9.0cm]{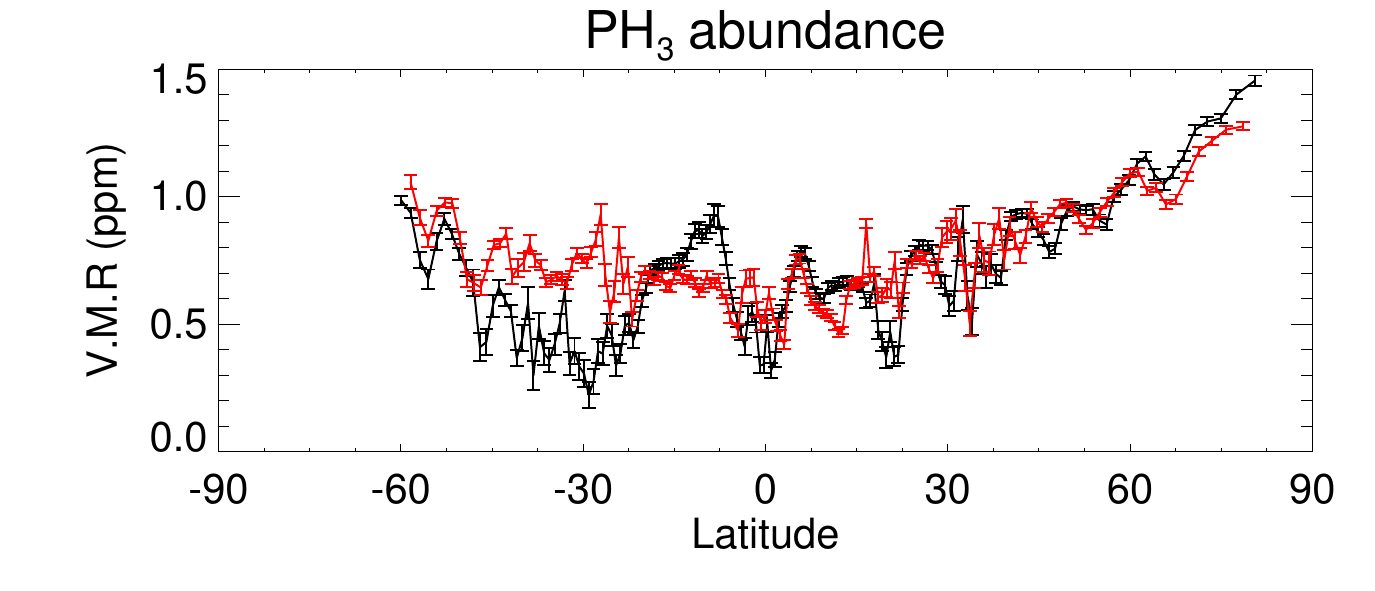}
\caption{PH\textsubscript{3} abundance as a function of latitude from the 12 November 2012 dataset (black) and the 1 January 2013 dataset (red). In both cases, a single cloud deck with $g=0.8$ is assumed and the AsH\textsubscript{3} abundance is allowed to vary. The two datasets produce consistent results.}
\label{fig:ph3_latitude_nov_jan}
\end{figure}

Figure~\ref{fig:ph3_latitude} is restricted to the data from 12 November 2012. However, as with AsH\textsubscript{3}, the 1 January 2013 dataset produces consistent results (see Figure~\ref{fig:ph3_latitude_nov_jan}), showing that the overall conclusion of an PH\textsubscript{3} enhancement at high latitudes is reproducible. 

\subsection{Summary}

In this section, we considered how the strength of a well-isolated PH\textsubscript{3} absorption line varied with latitude. We found that the depth of the absorption feature was greater for the high northern latitudes than for the equatorial latitudes. As with AsH\textsubscript{3}, we concluded that this was due to higher PH\textsubscript{3} abundances at high latitudes. We then performed a latitudinal retrieval, and found that this showed a minimum near the equator and an enhancement at high latitudes.

%%%%%%%%%%%%%%%%%%%%%%%%%%%%%%%%%%%%%%%%%%%%
%%%%%%%%%%%%%%%%%%%%%%%%%%%%%%%%%%%%%%%%%%%%

\section{Discussion}
\label{sec:discussion}

\subsection{Cloud structure and degeneracies}

In Sections~\ref{sec:asymmetry} and~\ref{sec:ch3d} we showed that there are significant degeneracies between Jupiter's tropospheric clouds and the gaseous composition. Firstly, Section~\ref{sec:asymmetry} showed that the retrieved abundances are sensitive to the cloud scattering properties; a high asymmetry parameter (more forward scattering, less reflected sunlight) leads to a low abundance, and a low asymmetry parameter (less forward scattering, more reflective sunlight) leads to a high abundance. The scattering properties also control the strength of the Fraunhofer lines in the spectrum, and by looking at the strength of these lines in the observations, we estimate that a value of $g=0.8$ provides the best fit to the data. However, lower values can also provide an adequate fit, meaning that the degeneracy is not entirely broken.

Secondly, Section~\ref{sec:ch3d} showed that the deep cloud structure is also degenerate with the retrieved gaseous abundances. Increasing the optical thickness of a deep cloud has a similar effect on the spectrum as decreasing the molecular abundance. An optically thin deep cloud and a low abundance can produce a similar spectral shape to an optically thick cloud and a high abundance. In the case of CH\textsubscript{3}D, this proves useful. The EqZ and SEB have different spectral shapes, yet we expect the CH\textsubscript{3}D abundance to be constant. As described in~\citet{bjoraker15}, the presence of a cloud at 5 bar that is optically thick in the zones and optically thin in the belts can solve this `problem'. These observations therefore provide some evidence for the existence of a deep cloud; based on the location, this cloud is likely to be composed of water ice. However, the degeneracy with the scattering properties limits the robustness of this conclusion. In the future, this could be improved by simultaneous modelling of the 5-\textmu m region and the near-infrared reflected sunlight spectrum. 

Both of these degeneracies complicate the retrievals of other molecular species. Since the cloud structure varies between belts and zones, these degeneracies especially limit our ability to come to reliable conclusions about belt-zone variability in gaseous abundances. This is further demonstrated in Section~\ref{sec:geh4}, where we showed that we were unable to distinguish between variability in GeH\textsubscript{4} abundance and variability in clouds. 

\subsection{Disequilibrium species}

\subsubsection{Abundances}

In Sections~\ref{sec:geh4}-\ref{sec:ph3}, we measured the abundances of three disequilibrium species in Jupiter's troposphere: GeH\textsubscript{4}, AsH\textsubscript{3} and PH\textsubscript{3}. For GeH\textsubscript{4}, we initially retrieved abundances that varied between 0.19 ppb (EqZ) and 0.58 ppb (SEB). By including a variable deep cloud, we were able to fit all spatial regions using a tropospheric abundance of 0.58 ppb. This is similar to previous results, which range from 0.45 ppb~\citep{bezard02} to 1.0 ppb~\citep{drossart82a}. This is significantly lower than the solar abundance of 6.7 ppb~\citep{grevesse07}, but this is unsurprising as GeS, rather than GeH\textsubscript{4}, is the most abundant Ge species~\citep{fegley94}.

Unlike GeH\textsubscript{4}, measurements of the AsH\textsubscript{3} absorption feature showed evidence for an enhancement at high latitudes. Using a one-cloud model, the retrieved abundance varies from as low as 0.01 ppb in parts of the equatorial region to 0.55 ppb at 80$^{\circ}$N. Previous measurements of the AsH\textsubscript{3} abundance have relied exclusively on data from the low latitudes of the planet, and have led to estimates of 0.2-0.3 ppb~\citep{bezard89,noll90,bezard02}. \citet{wang16} notes that AsH\textsubscript{3} is expected to be the primary As gas in the troposphere, and that these previous studies therefore suggest that the jovian abundance of As is lower than the solar value~\citep[a solar abundance would lead to a volume mixing ratio of 0.34 ppb,][]{grevesse07}. This is surprising, as other heavy elements show an enrichment relative to solar abundances~\citep[e.g.][]{taylor04}. Our results show that in some parts of the planet, the observed AsH\textsubscript{3} abundance is subsolar, while in other regions, AsH\textsubscript{3} shows an enrichment of up to 1.6 times the solar value. It is possible that this higher abundance is more representative of the deep abundance of As.

As with AsH\textsubscript{3}, the retrieved PH\textsubscript{3} abundance varies with latitude, ranging from 0.5 ppm in the equatorial regions to 1.4 ppm at 80$^{\circ}$N. These values are consistent with previous 5-\textmu m observations of PH\textsubscript{3}: 0.7 ppm~\citep{bjoraker86a}, 0.77 ppm~\citep{irwin98} and 0.76-0.90 ppm~\citep{giles15}. As noted in~\citet{giles15}, measurements at 5-\textmu m appear to consistently lead to lower retrieved abundances than measurements at 10-\textmu m~\citep[e.g.][]{fletcher09}; these differences are potentially due to errors in line data or cloud modelling uncertainties. Assuming a solar abundance of P would lead to a volume mixing ratio of 0.43 ppm~\citep{grevesse07}, so our analysis shows a variation from roughly solar levels to an enrichment of up to 3.3 times more the solar value, which is consistent with the observed enhancement of other heavy elements in Jupiter's atmosphere.

\subsubsection{Latitudinal profiles}

In Sections~\ref{sec:geh4}-\ref{sec:ph3}, we studied the latitudinal distributions of GeH\textsubscript{4}, AsH\textsubscript{3} and PH\textsubscript{3}. We found no evidence for spatial variability in the abundance of GeH\textsubscript{4}. This result for GeH\textsubscript{4} agrees with the previous study of~\citet{drossart82a} who found no evidence for variability in the 30$^{\circ}$S to 30$^{\circ}$N region, and our analysis shows that this remains true at higher latitudes. This is in contrast to the results for AsH\textsubscript{3} and PH\textsubscript{3}. While there is no evidence for any belt/zone variability, both of these species exhibit an enhancement towards the poles. An enhancement at high latitudes has previously been reported for PH\textsubscript{3}~\citep{drossart90,giles15}, but this is the first time that it has been seen in AsH\textsubscript{3}, as the only previous studies to compare different regions of the planet focussed exclusively on the equatorial latitudes~\citep{noll90}.

Our results contrast with the theoretical predictions from~\citet{wang16}. In their work, they use a diffusion-kinetics code to investigate how the abundances of disequilibrium species in Jupiter's troposphere depend on the vertical eddy diffusion coefficient, $K\textsubscript{eddy}$. They find that GeH\textsubscript{4} has a different behaviour to AsH\textsubscript{3} and PH\textsubscript{3}, which agrees with our observations. The GeH\textsubscript{4} abundance is found to be very sensitive to the eddy diffusion coefficient, with a higher value of $K\textsubscript{eddy}$ leading to a higher tropospheric abundance. In contrast, PH\textsubscript{3} and AsH\textsubscript{3} do not appear to have any dependence on $K\textsubscript{eddy}$ (for a physically plausible range of $K\textsubscript{eddy}$). These differences are due to the interplay between the quench level for the gas (the pressure at which the abundance is ``frozen in'') and the equilibrium vertical profile. If the quench level is deep in the regime where the gas is well-mixed, then a small change in $K\textsubscript{eddy}$ will not alter the observed abundance. If the quench level is in a regime where the equilibrium abundance is rapidly changing with altitude, then a small change in $K\textsubscript{eddy}$ can have a significant effect on the observed abundance. 

\citet{wang16} considers how abundances vary as a function of $K\textsubscript{eddy}$, while a previous study by the same authors~\citep{wang15} looks at how $K\textsubscript{eddy}$ varies with latitude. In~\citet{wang15}, they suggest that rotation plays an increasingly important role in suppressing turbulent convection at higher latitudes and that $K\textsubscript{eddy}$ should decrease towards the poles. This also agrees with earlier theoretical work by~\citet{flasar78}. Incorporating this finding, \citet{wang16} therefore predict that GeH\textsubscript{4} should decrease in abundance towards the poles, while AsH\textsubscript{3} and PH\textsubscript{3} should remain constant. In contrast, our observations show that GeH\textsubscript{4} is constant with latitude, while AsH\textsubscript{3} and PH\textsubscript{3} increase towards the poles. 

It is unclear what the cause of this discrepancy is. One possibility is that it is due to photolytic destruction, which was not considered in~\citet{wang16}. Many studies have concluded that ultraviolet photons are a significant contribution to the destruction of PH\textsubscript{3} in the upper troposphere~\citep[e.g.][]{strobel77,ferris84}. It has also been suggested as a destruction mechanism for GeH\textsubscript{4}~\citep{guillemin95} and could plausibly apply to AsH\textsubscript{3} too. Geometry effects mean that higher latitudes receive fewer UV photons from the sun, and the higher opacity of stratospheric aerosols in the polar regions also acts as a shield to UV light~\citep{zhang13}. Both of these effects could lead to a lower rate of photolytic destruction at high latitudes, and therefore a higher abundance. By increasing the abundances of all three species at high latitudes, the latitudinal profiles from~\citet{wang16} can be modified to match our observations. However, the reason that photolysis was neglected by~\citet{wang16} is that it is expected to primarily affect the upper troposphere (p$<$0.5 bar), not the deeper regions probed in the 5-\textmu m region~\citep{strobel77}.

Alternatively, it is possible that the vertical mixing strength does not in fact decrease towards the poles. \citet{wang15} and \citet{flasar78} focus on small-scale turbulent convection, and how this is affected by the increasing importance of rotation at high latitudes. While turbulent convection is inhibited towards the poles, it is possible that planetary-scale vertical motion increases. Solar heating is at a minimum at the poles, which reduces the static stability of the atmosphere and convection becomes uninhibited~\citep{irwin03}. However, ~\citet{wang16} showed that PH\textsubscript{3} and AsH\textsubscript{3} are not sensitive to the strength of vertical mixing, while GeH\textsubscript{4} is. Even if the trend reported in \citet{wang15} and \citet{flasar78} were to be reversed entirely, we would expect to see an increase of GeH\textsubscript{4} at the poles and a flat latitudinal distribution of AsH\textsubscript{3} and PH\textsubscript{3}, which is not what we observe. 

It is also possible that the sensitivities of the disequilibrium species to $K\textsubscript{eddy}$, as described in \citet{wang15}, are not robust. In several cases, the authors note a lack of available kinetic data which limit the confidence in the results. However, these diffusion-kinetics calculations do provide a useful explanation for the difference in behaviour between GeH\textsubscript{4} and AsH\textsubscript{3}/PH\textsubscript{3}. In addition, changes to the chemistry model cannot single-handedly account for the discrepancy between the theory and the observations. If AsH\textsubscript{3} and PH\textsubscript{3} were more sensitive to $K\textsubscript{eddy}$, they would be predicted to decrease, not increase, towards the poles, since $K\textsubscript{eddy}$ decreases at high latitudes~\citep{wang15}. 

Finally, there is the possibility that our observational conclusions are incorrect. We have shown that there are several degeneracies which complicate the retrievals of gaseous abundances in Jupiter's troposphere. Nevertheless, we are confident in the overall latitudinal trends that we report. The differences between the equatorial latitudes and the high latitudes are extremely apparent in the raw observations themselves, even before we begin to model the spectra (Figures~\ref{fig:ash3_fits} and~\ref{fig:ph3_fits}). 

%%%%%%%%%%%%%%%%%%%%%%%%%%%%%%%%%%%%%%%%%%%%
%%%%%%%%%%%%%%%%%%%%%%%%%%%%%%%%%%%%%%%%%%%%

\section{Conclusions}

The CRIRES instrument on the VLT was used to make high-resolution (R=96,000) observations of Jupiter in the 5-\textmu m window, where the planet's atmosphere is optically thin. This enabled us to spectrally resolve the line shapes of four minor species in Jupiter’s troposphere: CH\textsubscript{3}D, GeH\textsubscript{4}, AsH\textsubscript{3} and PH\textsubscript{3}. The slit was aligned north-south along Jupiter's central meridian, allowing us to search for latitudinal variability in these line shapes. The spectra were analysed using the NEMESIS radiative transfer code and retrieval algorithm. We make the following conclusions:

\begin{enumerate}

\item There are degeneracies between the cloud structure and scattering properties that complicate the retrievals of tropospheric gaseous abundances. These degeneracies can sometimes explain apparent belt-zone variations in composition. Any future studies into belt-zone variability must therefore carefully assess the degeneracies with the assumed scattering properties before any conclusions about spatial variations in gaseous abundances can be drawn. 

\item As seen in~\citet{bjoraker15}, there is some evidence for the existence of a highly variable cloud deck at $\sim$5 bar, but this conclusion is limited by degeneracies with cloud scattering properties.

\item We are able to fit all spatial regions with a GeH\textsubscript{4} abundance of 0.58 ppb, which is consistent with previous studies. We find that the AsH\textsubscript{3} abundance varies between 0.01 ppb and 0.55 ppb from equator to pole. This is the first time that supersolar abundances of AsH\textsubscript{3} have been measured in Jupiter's atmosphere. We measure a PH\textsubscript{3} abundance that varies between 0.5 and 1.5 ppm, which is consistent with previous measurements. 

\item We found that GeH\textsubscript{4} appears to have an abundance that is constant with latitude, while PH\textsubscript{3} and AsH\textsubscript{3} show evidence for an enhancement at high latitudes. This partially agrees with a theoretical study from~\citet{wang16} which suggests that GeH\textsubscript{4} should have a different latitudinal profile to PH\textsubscript{3} and AsH\textsubscript{3}. However, our observations disagree with the shapes of those profiles, as they predict that GeH\textsubscript{4} should decrease towards the poles while PH\textsubscript{3} and AsH\textsubscript{3} are flat. The explanation for this difference is not clear, but it could be due to the effect of photolytic destruction, increased global-scale convection at the poles, missing chemical kinetic data, or a combination thereof. 

\end{enumerate}

This study has demonstrated the utility of high-resolution 5-\textmu m observations in studying Jupiter's mid-tropospheric composition, revealing the latitudinal distributions of arsine and germane for the first time.  The CRIRES observations were limited to central-meridian scans of the planet.  Future work to explore Jupiter's deep atmosphere must utilise spatially-resolved spectral mapping to relate the dynamics of distinct atmospheric features to the distributions of these disequilibrium species.

%%%%%%%%%%%%%%%%%%%%%%%%%%%%%%%%%%%%%%%%%%%%
%%%%%%%%%%%%%%%%%%%%%%%%%%%%%%%%%%%%%%%%%%%%

\section*{Acknowledgements}

Giles was supported via a Royal Society studentship, and Fletcher was supported via a Royal Society Research Fellowship at the University of Leicester. Irwin acknowledges the support of the United Kingdom Science and Technology Facilities Council. This work is based on observations collected at the European Organisation for Astronomical Research in the Southern Hemisphere under ESO programme 090.C-0053(A).

\section*{References}

\bibliography{/Users/rohinigiles/Documents/Work/master.bib}

%\end{linenumbers}

\end{document}